\documentclass{ieeeaccess}
\usepackage{cite}
\usepackage{amsmath,amssymb,amsfonts}
\usepackage{algorithmic}
\usepackage{graphicx}
\usepackage{textcomp}
\usepackage{mathtools}
\usepackage{multirow}
\usepackage{url}
\usepackage{color}
\usepackage{setspace}
\usepackage{booktabs}
\usepackage{subfigure}
\usepackage{braket}
\usepackage{framed}

\def\BibTeX{{\rm B\kern-.05em{\sc i\kern-.025em b}\kern-.08em
    T\kern-.1667em\lower.7ex\hbox{E}\kern-.125emX}}

\begin{document}

\history{Date of publication xxxx 00, 0000, date of current version xxxx 00, 0000.}
\doi{10.1109/ACCESS.2023.3310875}

%\onecolumn
%\begin{framed}
%    \noindent
%    This work has been submitted to the IEEE for possible publication. Copyright may be transferred without notice, after which this version may no longer be accessible.%
%\end{framed}
%\clearpage
%\twocolumn

\title{Dynamical process of a bit-width reduced Ising model with simulated annealing}

\author{
\uppercase{Shuta Kikuchi}\authorrefmark{1},
\uppercase{Nozomu Togawa}\authorrefmark{2}, \IEEEmembership{Member, IEEE}, and
\uppercase{Shu Tanaka}\authorrefmark{1,3,4,5}
}

\address[1]{Department of Applied Physics and Physico-Informatics, Keio University, Kanagawa 223-8522, Japan}
\address[2]{Department of Computer Science and Communications Engineering, Waseda University, Tokyo 169-8555, Japan}
\address[3]{Human Biology-Microbiome-Quantum Research Center (WPI-Bio2Q), Keio University, Tokyo 108-8345, Japan}
\address[4]{Green Computing System Research Organization, Waseda University, Tokyo 162-0042, Japan}
\address[5]{International Research Frontiers Initiative, Tokyo Institute of Technology, Tokyo, 108-0023, Japan}

\markboth
{Shuta Kikuchi \headeretal: Dynamical process of a bit-width reduced Ising model with simulated annealing}
{Shuta Kikuchi \headeretal: Dynamical process of a bit-width reduced Ising model with simulated annealing}

\corresp{Corresponding author: Shuta Kikuchi (e-mail: skikuchi@keio.jp)}

\begin{abstract}
    Ising machines have attracted attention as efficient solvers for combinatorial optimization problems, which are formulated as ground-state (lowest-energy) search problems of the Ising model. Due to the limited bit-width of coefficients on Ising machines, the Ising model must be transformed into a bit-width reduced (BWR) Ising model. According to previous research, the bit-width reduction method, which adds auxiliary spins, ensures that the ground state of the BWR Ising model is theoretically the same as the Ising model before bit-width reduction (original Ising model). However, while the dynamical process is closely related to solution accuracy, how the BWR Ising model progresses towards the ground state remains to be elucidated. Therefore, we compared the dynamical processes of these models using simulated annealing (SA). Our findings reveal significant differences in the dynamical process across models. Analysis from the viewpoint of statistical mechanics found that the BWR Ising model has two characteristic properties: an effective temperature and a slow relaxation. These properties alter the temperature schedule and spin flip probability in the BWR Ising model, leading to differences in the dynamical process. Therefore, to obtain the same dynamical process as the original Ising model, we proposed SA parameters for the BWR Ising model. We demonstrated the proposed SA parameters using a square lattice Ising model, in which all coefficients were set uniformly to the same positive values or randomly. Our experimental evaluations demonstrated that the dynamical process of the BWR and original Ising model became closer.
\end{abstract}


\titlepgskip=-21pt

\maketitle

\section{Introduction}
\label{sec:introduction}

\subsection{Combinatorial optimization problem and Ising model}
\label{Combinatorial optimization problem and Ising model}

Combinatorial optimization problems find the optimal combination of decision variables to minimize or maximize the objective function for the given constraints. 
Typical examples include the traveling salesman problem, the Max-Cut problem, and the knapsack problem. 
Because such problems can be found in many real-world application domains, there is growing interest in developing techniques to find the optimal or quasi-optimal solution efficiently and accurately. 

Some combinatorial optimization problems can be formulated in a mathematically constructed model in statistical mechanics called an \textit{Ising model} or its equivalent model called a \textit{quadratic unconstrained binary optimization (QUBO) model}~\cite{lucas2014ising, Tanaka2017}.
The ground state of the Ising model corresponds to the optimal solution of the combinatorial optimization problem, where the ground state is referred to as the lowest-energy state. 

An Ising model is defined on an undirected graph $G=(V, E)$, where $V$ and $E$ are sets of vertices and edges, respectively.
The Ising model consists of spins, magnetic fields, and interactions. 
The Hamiltonian (or energy function) $H$ of the Ising model is defined by
\begin{align}
  H = - \sum_{i\in V}h_{i}\sigma_{i} - \sum_{(i,j)\in E} J_{ij}\sigma_{i}\sigma_{j} ,
  \label{eq:H}
\end{align}
where $\sigma_i$ is the spin on the vertex $i \in V$ and has a value of $(+1)$ or $(-1)$.
$h_{i}$ is the magnetic field on the vertex $i \in V$, and $J_{ij}$ is the interaction on the edge $(i, j) \in E$.
In this paper, we assume that interactions and magnetic fields are integer constants.

\subsection{Ising machine}

Approaches such as meta-heuristics and Ising machines have been developed to solve combinatorial optimization problems~\cite{Johnson2011, askarzadeh2016population, yamaguchi2016proposal, Yamaoka2016, Aramon2019, Inagaki2016, Goto2019, Maezawa2019, yamamoto2020statica, mohseni2022ising}. 
Ising machines have attracted attention as fast and high-precision solvers for combinatorial optimization problems. 
Ising machines specialize in searching for better solutions to combinatorial optimization problems formulated by an Ising model or a QUBO model.
Studies have applied Ising machines to various combinatorial optimization problems, including machine learning~\cite{neven2009, Amin2015, Amin2018, Daniel2018}, material design~\cite{Kitai2020,inoue2022towards, endo2022phase}, portfolio optimization~\cite{rosenberg2016solving, Tanahashi2019}, protein folding~\cite{Perdomo-Ortiz2012}, traffic optimization~\cite{Neukart2017, irie2019quantum, Bao2021-a,Bao2021-b,Mukasa2021}, quantum compiler~\cite{naito2023isaaq}, and black-box optimization~\cite{Kitai2020, izawa2022continuous, seki2022black}.

For an Ising machine to solve the problems formulated in the Ising model, the model must be mapped to the machine~\cite{Tanaka2020}. 
However, Ising machines are limited by their hardware specifications. 
For example, the number of spins corresponding to the problem size, the topology related to the connectivity between spins, and the \textit{bit-width} which is the imputable numerical range for coefficients of the interactions $J_{ij}$ and magnetic fields $h_i$. 
The specifications of the various Ising machines are summarized in~\cite{Oku2020,Kowalsky2021}. 
Although various approaches have been proposed in previous research to overcome the limitations due to the number of spins~\cite{qbsolv, karimi2017boosting, karimi2017effective, okada2019improving, irie2021hybrid, atobe2021hybrid, kikuchi2023hybrid} and the topology~\cite{choi2008minor, choi2011minor, cai2014practical, boothby2016fast, zaribafiyan2017systematic, oku2019, shirai2020guiding}, few studies have been devoted to overcoming the bit-width limitation.

\subsection{Motivation of this study}
The bit-width of a digital Ising machine, implemented by digital circuits such as Graphics Processing Unit (GPU), Field Programmable Gate Array (FPGA), or Application Specific Integrated Circuit (ASIC), is represented by an integer range of sign bits. Here, we assume that a bit-width of $n$-bits shows $[-(2^{n-1}-1), 2^{n-1}-1]$.

When the coefficients of the Ising model exceed the implemented bit-width of the Ising machine, they cannot be inputted into the Ising machine.
Thus, bit-width reduction methods such as the \textit{shift method} are used. 
The shift method divides by two until the coefficients of the Ising model fall within the target bit-width range~\cite{Oku2020}.
Although the shift method can naively reduce bit-width, the ground states of the bit-width reduced (BWR) Ising model may differ from that of the Ising model before bit-width reduction (original Ising model). 
Therefore, a new method to reduce bit-width by adding auxiliary spins is proposed~\cite{Oku2020}.
The proposed method guarantees that the ground states of the original Ising model and the BWR Ising model are theoretically consistent. 
However, the dynamical process of the BWR Ising model towards the ground state remains to be elucidated. 

This study analyzes the dynamical process of the BWR Ising model using simulated annealing (SA), which is the most fundamental algorithm for Ising machines implemented with digital circuits. 
The contributions of this study are as follows: 

\begin{itemize}
  \item The difference between the dynamical processes of the original Ising model and that of the BWR Ising model applying the proposed bit-width reduction method is elucidated. From a viewpoint of statistical mechanics, the BWR Ising model has two-characteristic properties: an effective temperature and a slow relaxation. These properties arise from the entropy effects of the auxiliary spins, which are not present in the original Ising model. 
  \item To obtain the same dynamical process of the original Ising model, we propose the setting parameters of the BWR Ising model for SA in which the temperature schedule and inner loop are modified. The effectiveness of the proposed SA parameters is evaluated using a dynamical process with square lattice random Ising models. The dynamical process of the BWR Ising model is equivalent to the original Ising model. 
\end{itemize}

The rest of this paper is organized as follows. 
Section~\ref{sec:method} introduces the bit-width reduction method. 
Section~\ref{sec:simulation} investigates the difference in dynamical processes between the original Ising model, which has known properties, and the BWR Ising model to clarify the dynamical properties of the BWR Ising model. 
Section~\ref{sec:analysis} presents the statistical mechanics analysis results of the BWR Ising model. 
Section~\ref{sec:proposed_parameter} proposes BWR Ising model parameters for SA. 
The experimental evaluations demonstrate that the dynamical processes of the BWR Ising model and the original Ising model are almost the same. 
Sections~\ref{sec:random_ising} and ~\ref{sec:Discussion} demonstrate and discuss the numerical results, respectively.
Section~\ref{sec:conclusion} concludes with a summary of our study and future research directions.
The Appendices provide supplemental information for the derivation of the statistical mechanics analysis for the BWR Ising model (Appendix~\ref{sec:appendixA}),  the effectiveness of the proposed SA parameter for large-size square lattice systems (Appendix~\ref{sec:appendixB}) and various temperature schedules (Appendix~\ref{sec:appendixC}). 

\section{method}
\label{sec:method}

\subsection{Bit-width reduction method}
\label{subsec:bit-width reduction method}
\begin{figure}[t]
  \centering
  \subfigure[]{
    \includegraphics[clip,width=0.45\linewidth]{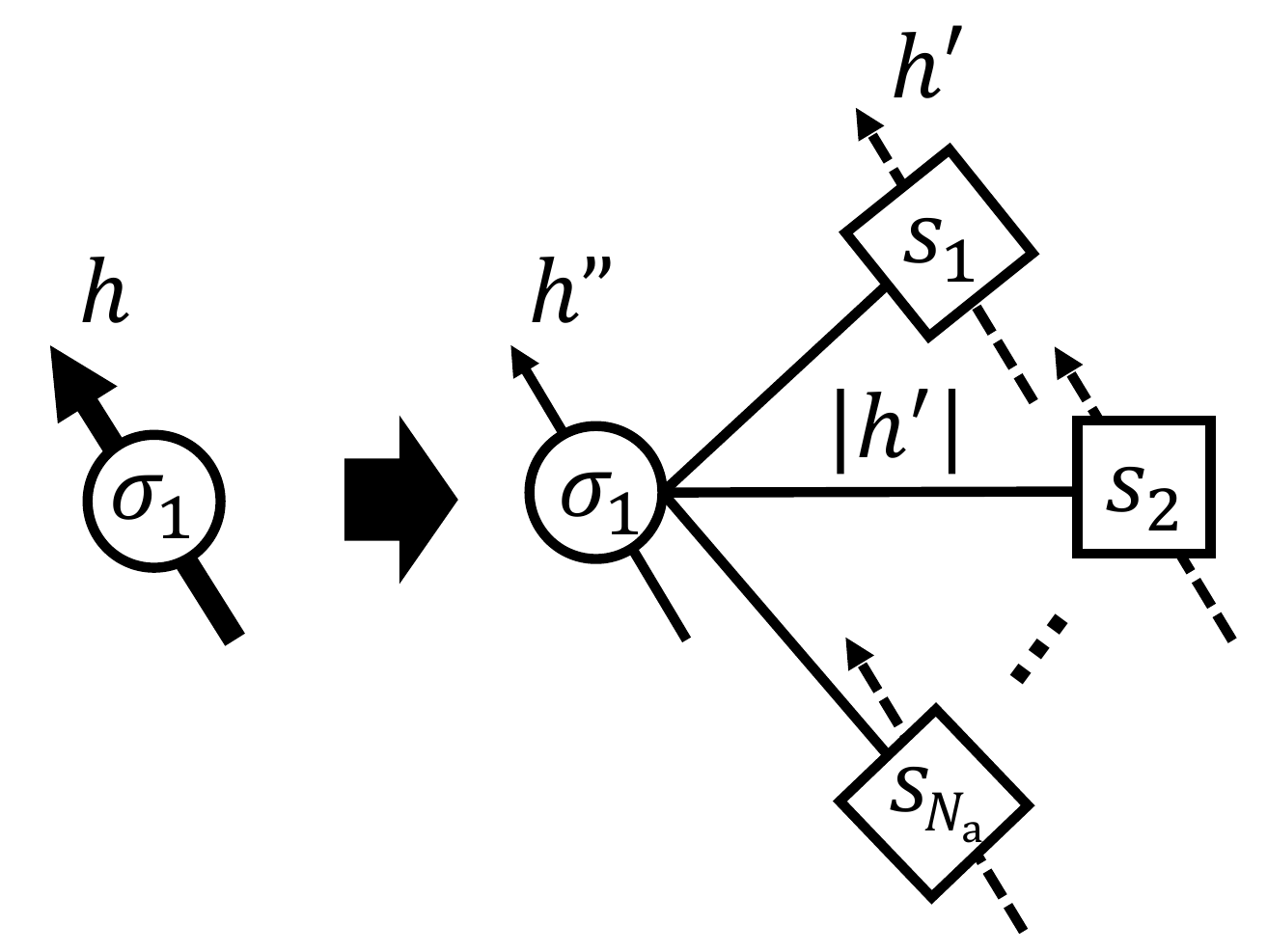}
    \label{fig:red_method_mag}
  }
  \subfigure[]{
    \includegraphics[clip,width=0.45\linewidth]{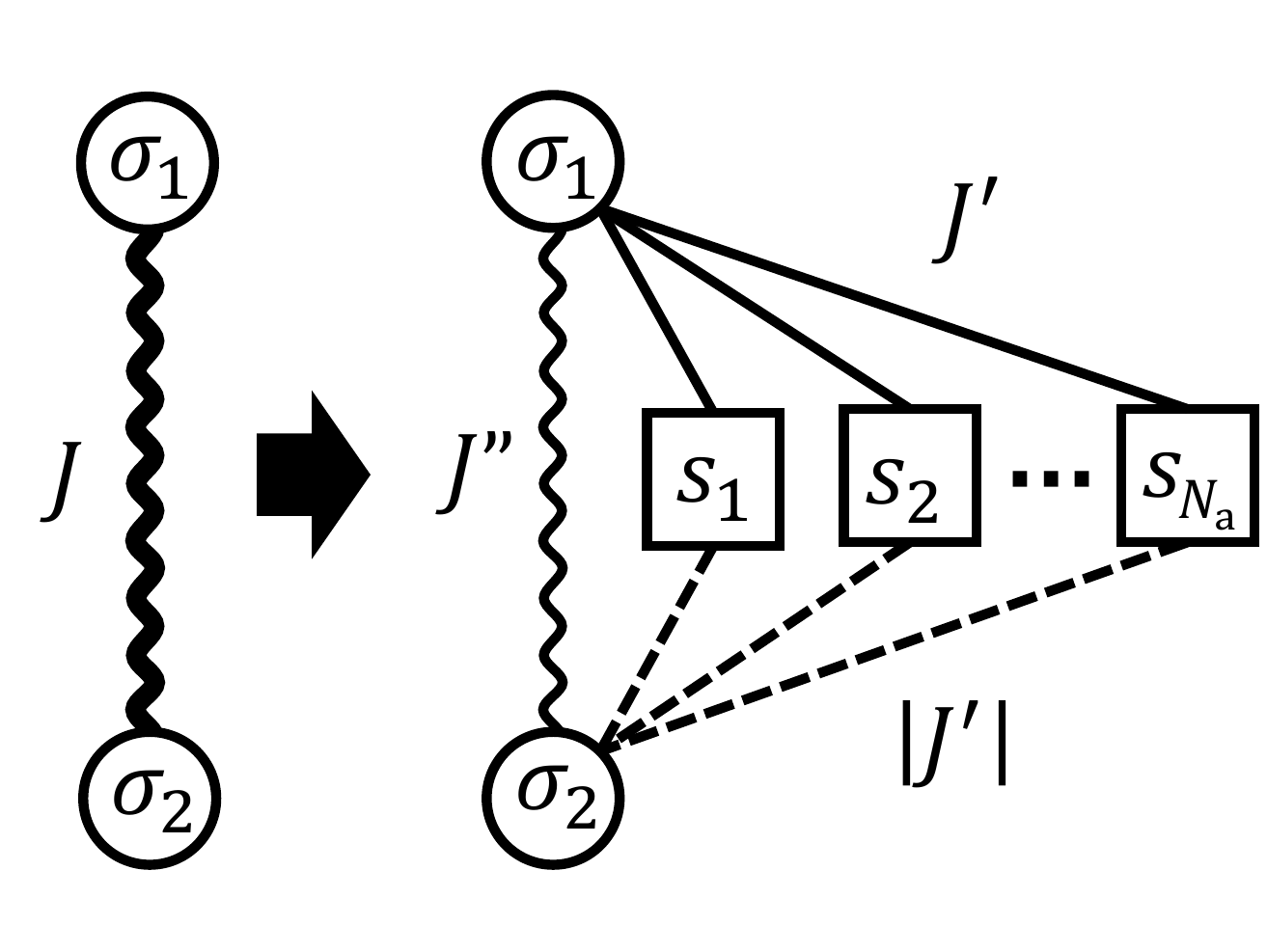}
    \label{fig:red_method_int}
  }
  \caption{
  Examples of bit-width reduction using the proposed method. Arrows, lines, circles, and squares represent the magnetic fields,  interactions, system spins, and auxiliary spins, respectively. (a) Bit-width reduction process of the magnetic fields. Thick solid, dotted, and thin solid arrows denote $h$, $h'$, and $h''$, respectively. Lines between system spin and auxiliary spins denote $|h'|$. (b) Bit-width reduction process of the interactions. Thick wavy, solid, dotted, and thin wavy lines denote $J$, $J'$, $|J'|$, and $J''$, respectively.}
  \label{fig:red_method}
\end{figure}

A previous study reported a bit-width reduction method, which added auxiliary spins~\cite{Oku2020}.
Herein the proposed bit-width reduction method modifies the previous method for statistical mechanics analysis.
Fig.~\ref{fig:red_method} depicts the bit-width reduction processes using the proposed method.
Although the previous and proposed methods have different coefficient assignments after bit-width reduction, the substantive static properties are the same.
Even with the proposed method, the ground state of the BWR Ising model and the original Ising model remain theoretically consistent. 

This subsection details the proposed method for bit-width reduction. 
The upper and lower limits of the target $n$-bits coefficients are $c^{\textrm{upper}}_n (=2^{n-1}-1)$ and $c^{\textrm{lower}}_n (=-(2^{n-1}-1))$, respectively.
First, we describe the method to reduce the bit-width of the magnetic fields in the original Ising model. 
We assume that the original Ising model includes a magnetic field $h$ acting on a spin $\sigma_1$ (Fig.~\ref{fig:red_method_mag}, left). 
The spin consisting of the original Ising model is called the ``system spin'' such as $\sigma_1$.
Applying the proposed method gives the BWR Ising model (Fig.~\ref{fig:red_method_mag}, right). 
The spins $s_i$ added by the proposed method are called ``auxiliary spins.''
The bit-width of the magnetic field coefficient $h$ is reduced to $n$-bits as follows: 

\begin{description}
   \item[Step 1:] Let $h''$ be the new magnetic field of $\sigma_1$, satisfying 
   \begin{align}
        h''=
        \begin{cases}
           h - N_\textrm{a} \times c^{\textrm{upper}}_n, ~0 < h''\leqq c^{\textrm{upper}}_n  & (h > 0) \\
           h - N_\textrm{a} \times c^{\textrm{lower}}_n, ~c^{\textrm{lower}}_n \leqq h'' < 0 & (h < 0)
        \end{cases}
        .
   \end{align}
   \item[Step 2:]Add $N_\textrm{a}$ auxiliary spins $s_i$ ($i = 1, 2, ..., N_\textrm{a}$). Let $h'$ be the magnetic fields of all auxiliary spins $s_i$, where 
   \begin{align}
       h' = 
       \begin{cases}
          c^{\textrm{upper}}_n & (h > 0) \\
          c^{\textrm{lower}}_n & (h < 0)
       \end{cases}
       .
   \end{align}
   \item[Step 3:]Introduce the interactions $|h'|$ between $\sigma_1$ and all auxiliary spins $s_i$.
\end{description}

Fig.~\ref{fig:red_method_int} shows the scheme to reduce the bit-width of the interactions.
We assume that two spins $\sigma_1$ and $\sigma_2$ are connected by the interaction $J$.
The bit-width of the interaction coefficient $J$ is reduced to $n$-bits as follows: 

\begin{description}
   \item[Step 1:] Let $J''$ be the new interaction between $\sigma_1$ and $\sigma_2$, satisfying 
   \begin{align}
       J'' = 
       \begin{cases}
          J - N_\textrm{a} \times c^{\textrm{upper}}_n, ~0 < J''\leqq c^{\textrm{upper}}_n & (J > 0) \\
          J - N_\textrm{a} \times c^{\textrm{lower}}_n, ~c^{\textrm{lower}}_n \leqq J'' < 0 & (J < 0)
       \end{cases}
       .
   \end{align}
   \item[Step 2:]Add $N_\textrm{a}$ auxiliary spins $s_i$ ($i = 1, 2, ..., N_\textrm{a}$). Let $J'$ be the interactions between $\sigma_1$ all auxiliary spins $s_i$, where 
   \begin{align}
       J' = 
       \begin{cases}
          c^{\textrm{upper}}_n & (J > 0) \\
          c^{\textrm{lower}}_n & (J < 0)
       \end{cases}
       .
   \end{align}
   \item[Step 3:]Introduce the interactions $|J'|$ between $\sigma_2$ and all auxiliary spins $s_i$.
\end{description}

\begin{figure}[t]
  \centering
  \subfigure[]{
    \includegraphics[clip,width=0.45\columnwidth]{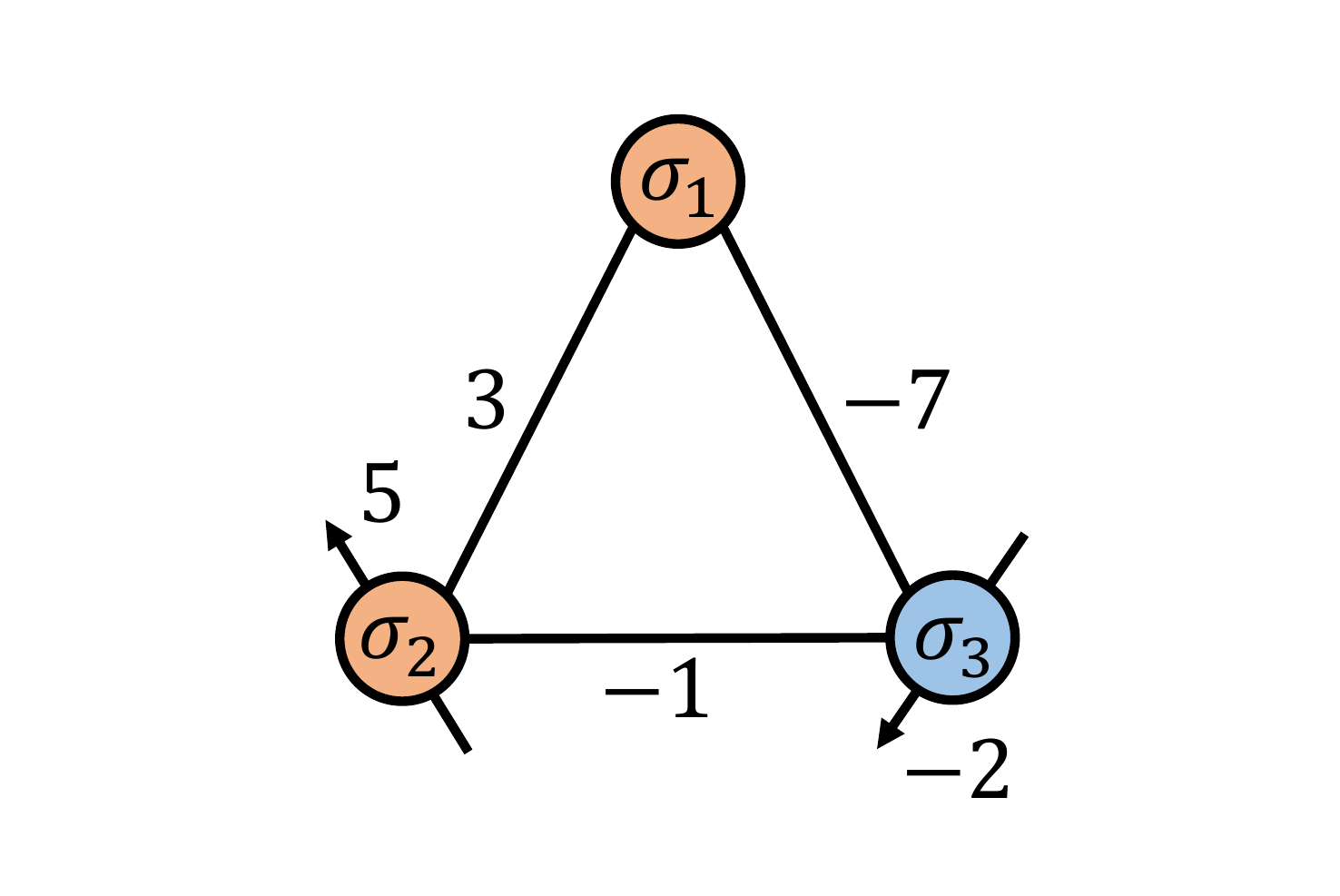}
    \label{fig:exa_4}
  }
  \subfigure[]{
    \includegraphics[clip,width=0.45\columnwidth]{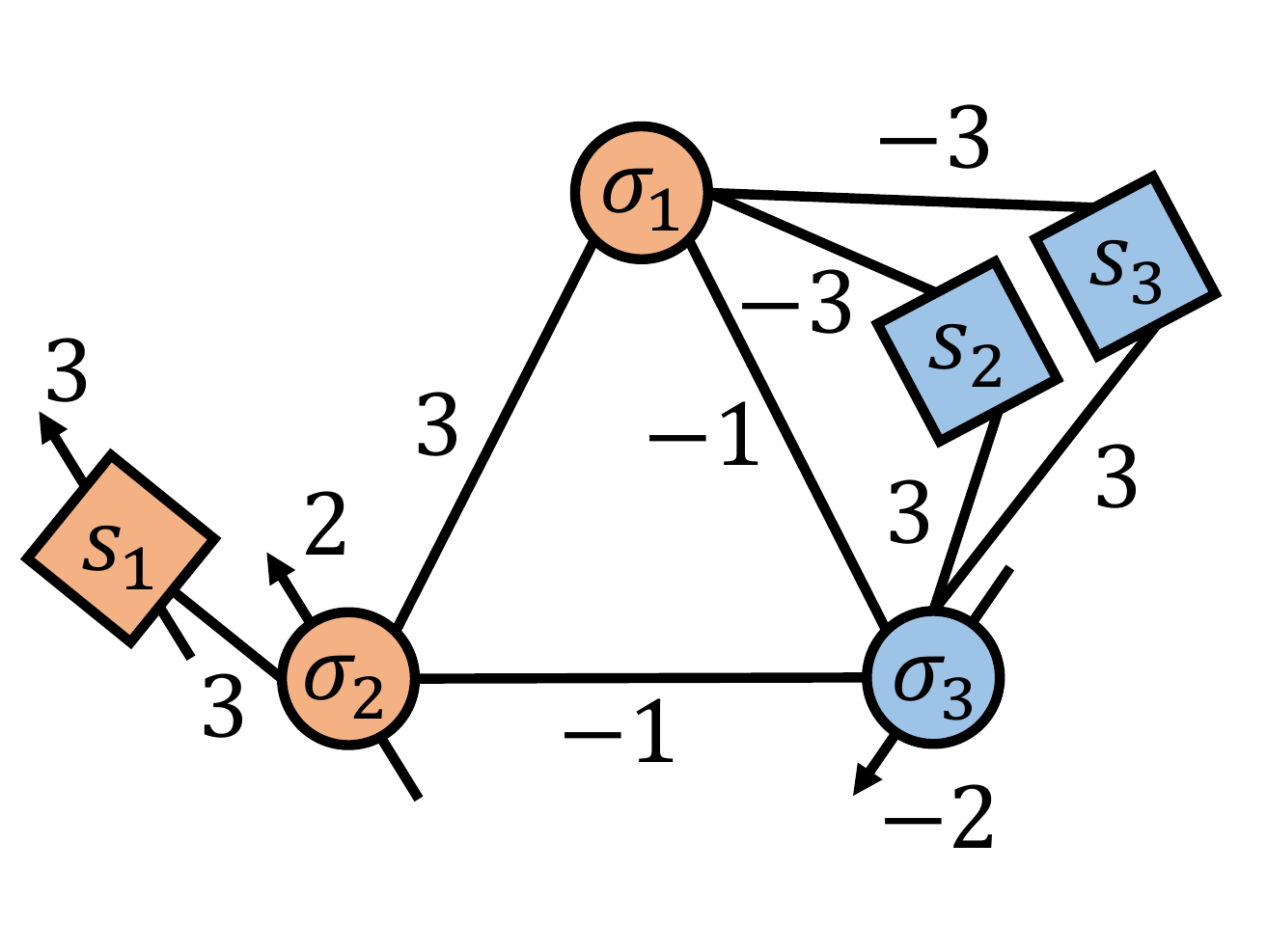}
    \label{fig:exa_3}
  }
  \caption{
  Example of bit-width reduction. Solid arrows, solid lines, circles, and squares show the magnetic fields, their interactions, system spins, and auxiliary spins, respectively. Orange and blue spins denote $+1$ and $-1$, respectively. (a) Original Ising model ($4$-bits). (b) BWR Ising model after applying the proposed method ($3$-bits).}
  \label{fig:exa_bit}
\end{figure}

Figs.~\ref{fig:exa_4} and~\ref{fig:exa_3} show examples of the original Ising model and the BWR Ising model obtained by applying the proposed method, respectively.
In this case, the bit-width of the coefficient is reduced from $4$-bits to $3$-bits.
The ground state of the original Ising model is ($\sigma_1$, $\sigma_2$, $\sigma_3$)$=$($+1$, $+1$, $-1$).
Similarly, the ground state of the BWR Ising model is ($\sigma_1$, $\sigma_2$, $\sigma_3$, $s_1$, $s_2$, $s_3$)$=$($+1$, $+1$, $-1$, $+1$, $-1$, $-1$). 
Focusing on the system spins, the ground states of the original Ising model and the BWR Ising model are clearly consistent. 

\subsection{SA}
\label{subsec:SA}

SA is a meta-heuristic algorithm with a wide range of applications~\cite{kirkpatrick1983optimization,johnson1989optimization,johnson1991optimization, isakov2015optimised}. 
During SA for $N$ spins Ising model given by (\ref{eq:H}), the following procedures are performed: 

\begin{description}
   \item[Step 1:]Prepare a random initial spin state. 
   \item[Step 2:]Set the initial temperature sufficiently high for the Hamiltonian. 
   \item[Step 3:]Choose one spin from $N$ spins randomly.
   \item[Step 4:]Flip the chosen spin according to the transition probability $W(\Delta{E}, T)$, which depends on the temperature $T$ and the energy difference $\Delta E$. Energy difference $\Delta E$ is defined by $\Delta{E}=H_\textrm{candidate}-H_\textrm{current}$, where $H_\textrm{candidate}$ is the energy of the candidate state in which the chosen spin is flipped and $H_\textrm{current}$ is the energy of the current state. Here, the transition probability, called the heat-bath method, is used and is expressed as $W(\Delta{E}, T)=[1 + \exp(\Delta{E}/T)]^{-1}$. 
   \item[Step 5:]Repeat Steps $3$–$4$ ``inner loop'' times. The inner loop is typically set to the number of spins $N$, which is called one Monte Carlo Step (MCS). 
   \item[Step 6:]Decrease the temperature $T$ and return to Step $3$. 
   \item[Step 7:]Repeat Step $6$, ``outer loop'' times. 
\end{description}

The Geman--Geman theorem guarantees that the ground state is ideally obtained in SA when the temperature decreases sufficiently slow\cite{Geman1984}.
Notice that the ground state may not be available and a lower-energy state (not the ground state) may be obtained in a realistic time.

\section{Dynamical process with SA}
\label{sec:simulation}

In this study, the dynamical properties of the BWR Ising model were clarified by comparing the dynamical process of the original Ising model to that of the BWR Ising model under SA.
We employed an Ising model on square $L \times L$ systems with periodic boundary conditions~\cite{instances}. 
Here, we set $L=30$ and the coefficients of magnetic fields and interactions to 7, that is, $h_i=7$ for all $i$ and $J_{ij}=7$ for all nearest-neighbor pairs on square lattice $i,j$ in~\eqref{eq:H}.
The properties of the Ising model are well-known.
In the ground state of the original Ising model, all spins take $+1$, with an internal energy per spin (i.e., energy density, $H/L^2$)  of $-21$.
In this demonstration, the bit-width of the coefficient is reduced from $4$-bits to $3$- or $2$-bits. 

Table~\ref{table:parameters_simulation} shows the SA parameters. 
The initial temperature $T_\textrm{initial}$ is set sufficiently high to permit the transition between arbitrary states at the beginning of SA. 
The temperature schedule is set to the power-law decay for every outer loop, which is given by $T(t)=T_\textrm{initial} \times r^t$, where $r$ is the cooling rate and $t$ is the $t$-th outer loop.
The outer loop and cooling rate $r$ is set to 100 ($t = 0 - 99$) and 0.97, respectively. 
Using these values, the final temperature of SA becomes $2.451$, which is sufficiently low on the energy scale of the original Ising model. 
The inner loop is set to the number of spins in the Ising model ($1$ MCS). 
\begin{table}[t]
  \centering
  \caption{SA parameters.}
  \label{table:parameters_simulation}
  \begin{tabular}[t]{ll} \toprule
    Parameter & Value \\ \midrule
    Initial state & Random  \\
    \shortstack[l]{Initial temperature \\($T_\textrm{initial}$)} & \raisebox{0.5em}{ 50}  \\
    Cooling rate ($r$) & 0.97  \\
    Outer loop & 100  \\
    \raisebox{0.5em}{Inner loop} & \shortstack[l]{Number of spins  \\ (1 MCS)} \\ \bottomrule
  \end{tabular}
\end{table}

\begin{figure}[t]
  \centering
  \includegraphics[clip,width=0.9\linewidth]{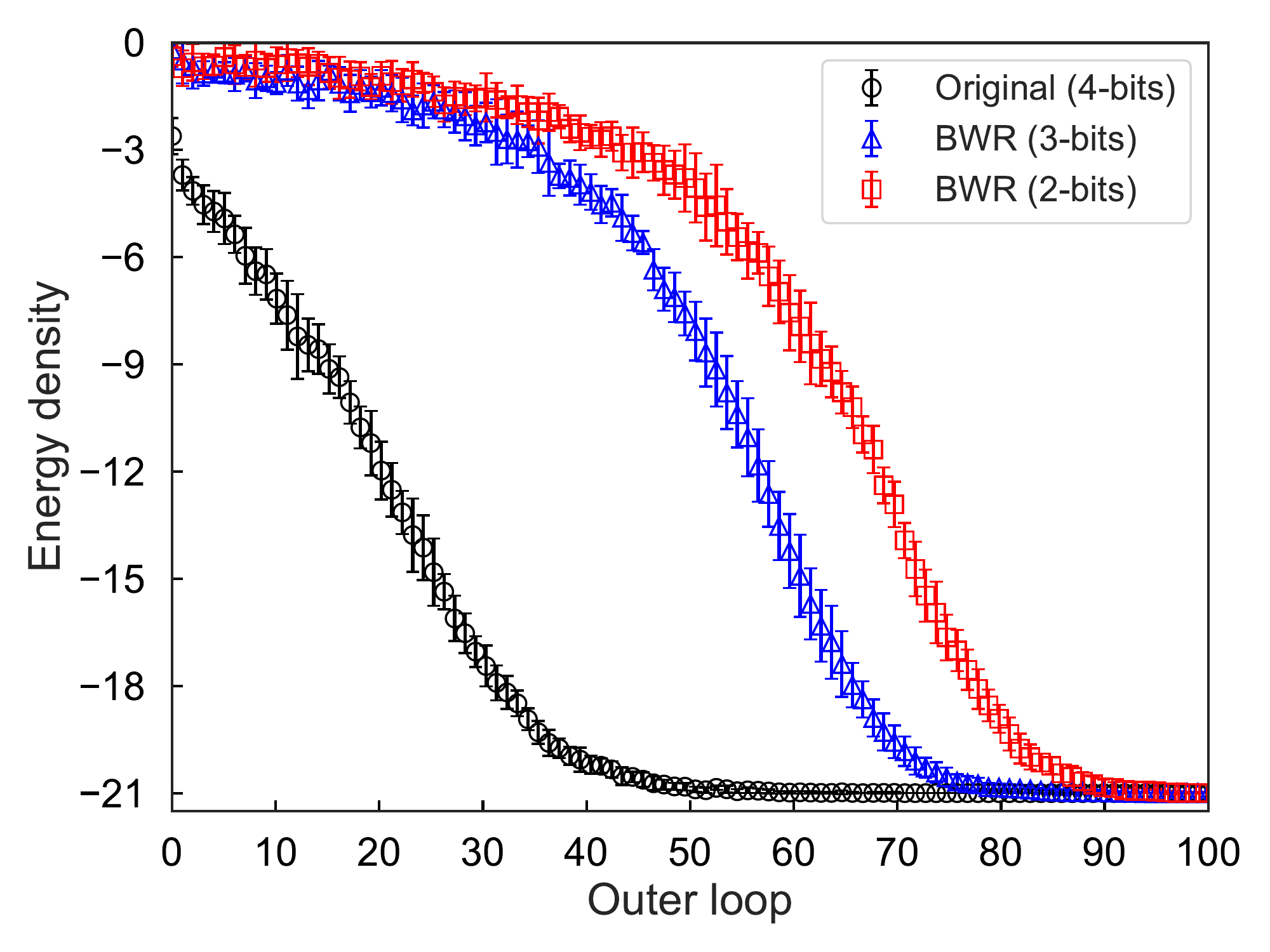}
  \caption{Dynamical processes of the original Ising model and the BWR Ising model. The BWR Ising model ($2$-bits), the BWR Ising model ($3$-bits) and the original Ising model ($4$-bits) are denoted by red squares, blue triangles and black circles. Every plot is an average of ten runs. The error bars are standard deviations.}
  \label{fig:simulateion}
\end{figure}

Fig.~\ref{fig:simulateion} shows the experimental results. 
The energy density of both the original and BWR Ising models was calculated using the number of system spins (i.e., $L^2$).
The data were obtained from average and standard deviation of energy density for ten simulations of SA.
Although each BWR Ising model eventually yields the ground state, the dynamical process significantly differs from that of the original Ising model. 
Even at the steps of the outer loop, where the energy density decreases in the original Ising model, it did not decrease in the BWR Ising model.
Similar results were obtained even for large-size square lattice system ($L=40, 50$) in Appendix~\ref{sec:appendixB}.

\section{Analysis of the bit-width reduced Ising model}
\label{sec:analysis}

To investigate the difference in the dynamical processes between the original Ising model and the BWR Ising model, we analyzed the BWR Ising model from the viewpoint of the microscopic mechanism: effective temperature and slow relaxation. 

\subsection{Effective Temperature}
\label{subsec:Teff}

Previous studies in statistical mechanics employed an Ising model with a structure similar to the BWR Ising model (Fig.~\ref{fig:red_method_int})~\cite{Miyashita2007, Tanaka2009, tanaka2010nonmonotonic}.
It indicated that the dynamical processes of the correlation function between the system spins $\sigma_1$ and $\sigma_2$ $\lparen\langle\sigma_1\sigma_2\rangle\rparen$ of the temperature differ from the Ising model with and without the auxiliary spins~\cite{Miyashita2007}.
Here, $\langle \cdot \rangle$ represents the expectation value.
Therefore, we analyzed the BWR Ising model by referencing the previous studies. 

\subsubsection{Magnetic fields}
\label{subsec:magnetic_fields}
First, we considered the case where the bit-width of the magnetic fields is reduced by adding $N_\textrm{a}$ auxiliary spins (Fig.~\ref{fig:red_method_mag}).
The effective magnetic field $L_\textrm{eff}$ at temperature $T$ is defined as (see Appendix~\ref{sec:appendixA} for a detailed derivation)
\begin{align}
  \sum_{s_i=\pm1}e^{-\beta{H}}=A(T)e^{L_\mathrm{eff}\sigma_1},
  \label{eq:Leff_def}
\end{align}
where $\beta=1/T$ and $A(T)$ is an analytic function of $T$. 

When $h>0$, the Hamiltonian of the BWR Ising model depicted on the right of Fig.~\ref{fig:red_method_mag} is given by
\begin{align}
  H=-h''\sigma_1-h'\left(\sum^{N_\mathrm{a}}_{i=1}s_i\right)(1+\sigma_1),
  \label{eq:mg_posi_hamiltonian}
\end{align}
and the effective magnetic field of the system spin is obtained as 
\begin{align}
  L_\mathrm{eff}=\frac{h''}{T}+\frac{N_\mathrm{a}}{2}\log\left[\cosh\left(\frac{2h'}{T}\right)\right].
  \label{eq:mag_posi_Leff}
\end{align}

The effective temperature $T_\textrm{eff}=h/L_\textrm{eff}$ is given by
\begin{align}
  T_\mathrm{eff}=\frac{h}{\dfrac{h''}{T}+\dfrac{N_\mathrm{a}}{2}\log\left[\cosh\left(\dfrac{2h'}{T}\right)\right]}.
  \label{eq:mag_posi_Teff}
\end{align}

When $h<0$, the Hamiltonian depicted on the right of Fig.~\ref{fig:red_method_mag} is given by
\begin{align}
  H=-h''\sigma_1-h'\left(\sum^{N_\mathrm{a}}_{i=1}s_i\right)(1-\sigma_1),
  \label{eq:mag_nega_hamiltonian}
\end{align}
and the effective magnetic field is obtained as
\begin{align}
  L_\mathrm{eff}=\frac{h''}{T}-\frac{N_\mathrm{a}}{2}\log\left[\cosh\left(\frac{2h'}{T}\right)\right].
  \label{eq:mag_nega_Leff}
\end{align}

The effective temperature is given by
\begin{align}
  T_\mathrm{eff}=\frac{h}{\dfrac{h''}{T}-\dfrac{N_\mathrm{a}}{2}\log\left[\cosh\left(\dfrac{2h'}{T}\right)\right]}.
  \label{eq:mag_nega_Teff}
\end{align}

\subsubsection{Interactions}
\label{subsec:interactions}

Next, we considered the case where the bit-width of the interactions is reduced by adding $N_\textrm{a}$ auxiliary spins (Fig.~\ref{fig:red_method_int}).

The effective interaction $K_\textrm{eff}$ at temperature $T$ is defined as (see Appendix~\ref{sec:appendixA} for a detailed derivation)
\begin{align}
  \sum_{s_i=\pm1}e^{-\beta{H}}=A(T)e^{K_\mathrm{eff}\sigma_1\sigma_2}.
  \label{eq:Keff_def}
\end{align}

When $J>0$, the Hamiltonian of the BWR Ising model depicted on the right of Fig.~\ref{fig:red_method_int} is given by
\begin{align}
  H=-J''\sigma_1\sigma_2-J'\left(\sum^{N_\mathrm{a}}_{i=1}s_i\right)(\sigma_1+\sigma_2),
  \label{eq:int_posi_hamiltonian}
\end{align}
and the effective interaction between $\sigma_1$ and $\sigma_2$ is obtained as
\begin{align}
  K_\mathrm{eff}=\frac{J''}{T}+\frac{N_\mathrm{a}}{2}\log\left[\cosh\left(\frac{2J'}{T}\right)\right].
  \label{eq:int_posi_Keff}
\end{align}

The effective temperature $T_\textrm{eff}=h/K_\textrm{eff}$ is given by
\begin{align}
  T_\mathrm{eff}=\frac{J}{\dfrac{J''}{T}+\dfrac{N_\mathrm{a}}{2}\log\left[\cosh\left(\dfrac{2J'}{T}\right)\right]}.
  \label{eq:int_posi_Teff}
\end{align}

When $J<0$, the Hamiltonian depicted on the right of Fig.~\ref{fig:red_method_int} is given by
\begin{align}
  H=-J''\sigma_1\sigma_2-J'\left(\sum^{N_\mathrm{a}}_{i=1}s_i\right)(\sigma_1-\sigma_2),
  \label{eq:int_nega_hamiltonian}
\end{align}
and the effective interaction is obtained as
\begin{align}
  K_\mathrm{eff}=\frac{J''}{T}-\frac{N_\mathrm{a}}{2}\log\left[\cosh\left(\frac{2J'}{T}\right)\right].
  \label{eq:int_nega_Keff}
\end{align}

The effective temperature is given by
\begin{align}
  T_\mathrm{eff}=\frac{J}{\dfrac{J''}{T}-\dfrac{N_\mathrm{a}}{2}\log\left[\cosh\left(\dfrac{2J'}{T}\right)\right]}.
  \label{eq:int_nega_Teff}
\end{align}

\begin{figure}[t]
  \centering
  \includegraphics[clip,width=0.9\linewidth]{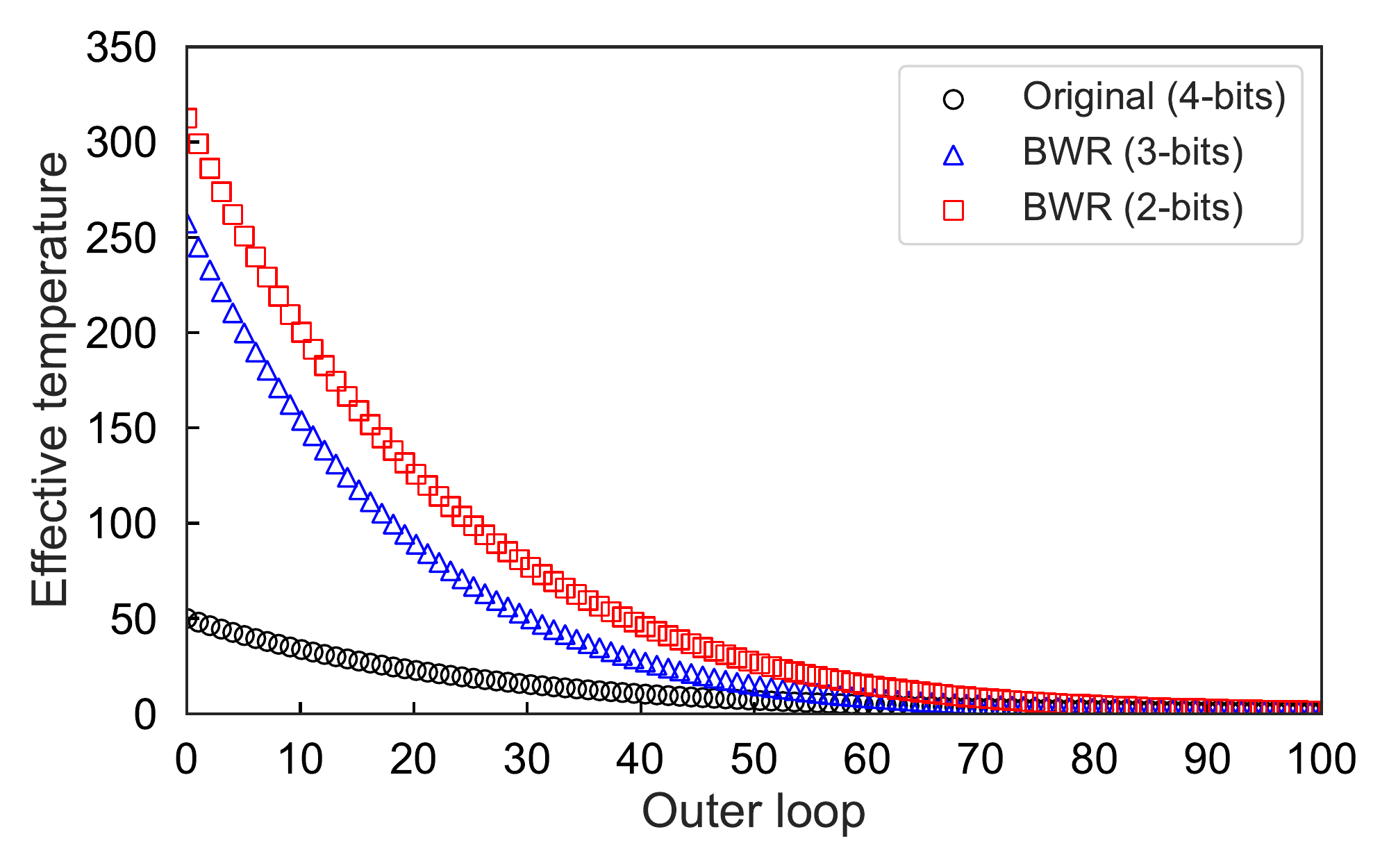}
  \caption{
  Effective temperature schedule of the original Ising model ($J$ or $h = 7$) and the BWR Ising model. Red squares, blue triangles, and black circles denote the BWR Ising model ($2$-bits), BWR Ising model ($3$-bits), and original Ising model ($4$-bits), respectively.}
  \label{fig:Teff_comp}
\end{figure}

Equations~(\ref{eq:mag_posi_Teff}), ~(\ref{eq:mag_nega_Teff}), ~(\ref{eq:int_posi_Teff}), and~(\ref{eq:int_nega_Teff}) indicate that the effective temperature $T_\textrm{eff}$ differs from the temperature $T$ added to the Ising model. 
Fig.~\ref{fig:Teff_comp} shows the effective temperature in the BWR Ising model, which was determined by comparing the temperature $T$ used for SA in the previous section and $T_\textrm{eff}$.
Since $J = h = 7$ is assumed, we set $N_\textrm{a} = 2$, $J'' = 1$, and $J' = 3$ for the calculation to reduce the bit-width to $3$-bits. 
To reduce the bit-width of coefficients to $2$-bits, we set $N_\textrm{a} = 6$ and $J'' = J' = 1$.
The temperature schedule of $T_\textrm{eff}$ rapidly decreases at a temperature above that of $T$.
This suggests that the discrepancy between the temperature $T$ and  $T_\textrm{eff}$ affects the dynamical process.

\subsection{Characteristic time scale}
\label{subsec:entropic_effect}

The previous study reported that a slow relaxation occurs in the lattice of frustrated systems with decorated spins~\cite{Tanaka2009}. 
This phenomenon is called ``entropic slowing down'' and is due to the degrees of freedom distribution of the decoration spins. 
The decorated lattice system has a similar structure to the BWR Ising model when applying the proposed method. 
Therefore, we assumed that an entropic slowing down appears in the BWR Ising model, and this phenomenon influences the dynamical processes. 
Following~\cite{Tanaka2009}, we determined the number of states when the local configuration of the system spins is fixed. 

\begin{figure}[b]
  \centering
  \subfigure[]{
    \includegraphics[clip,width=0.4\linewidth]{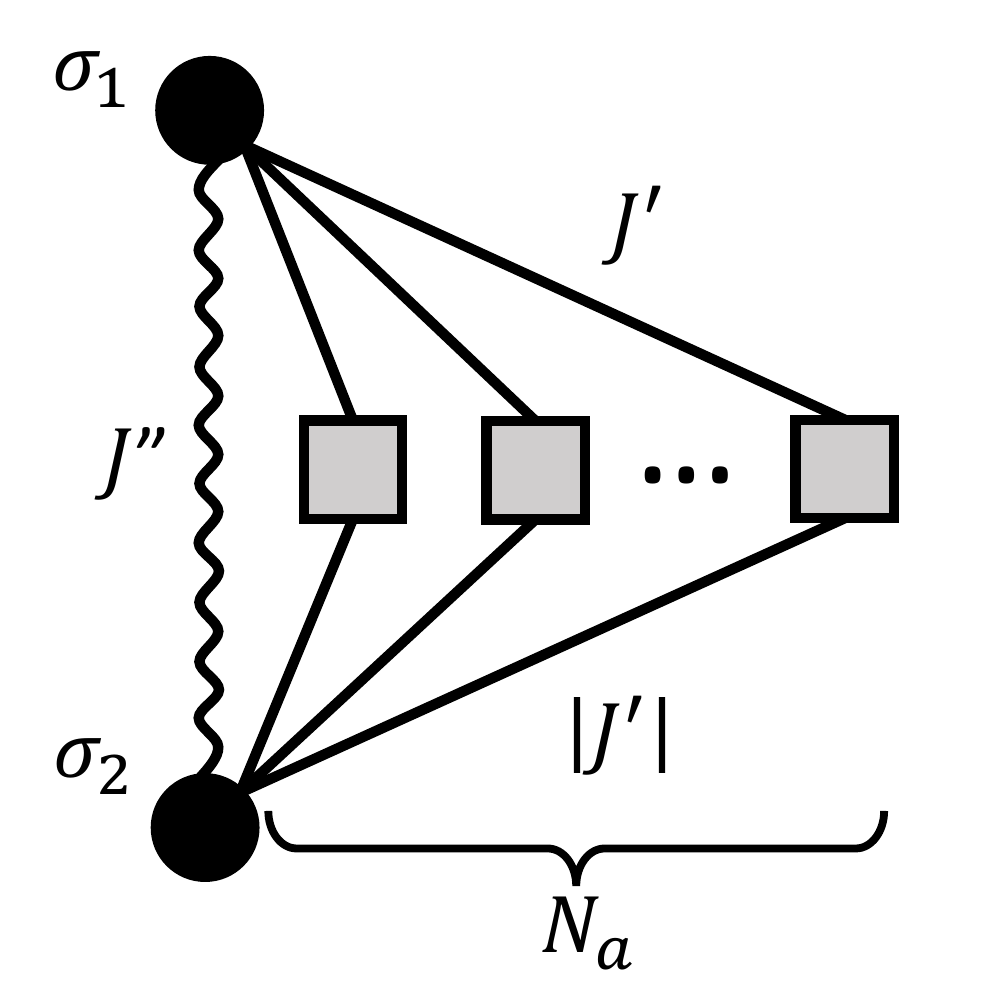}
    \label{fig:posi_para}
  }
  \subfigure[]{
    \includegraphics[clip,width=0.4\linewidth]{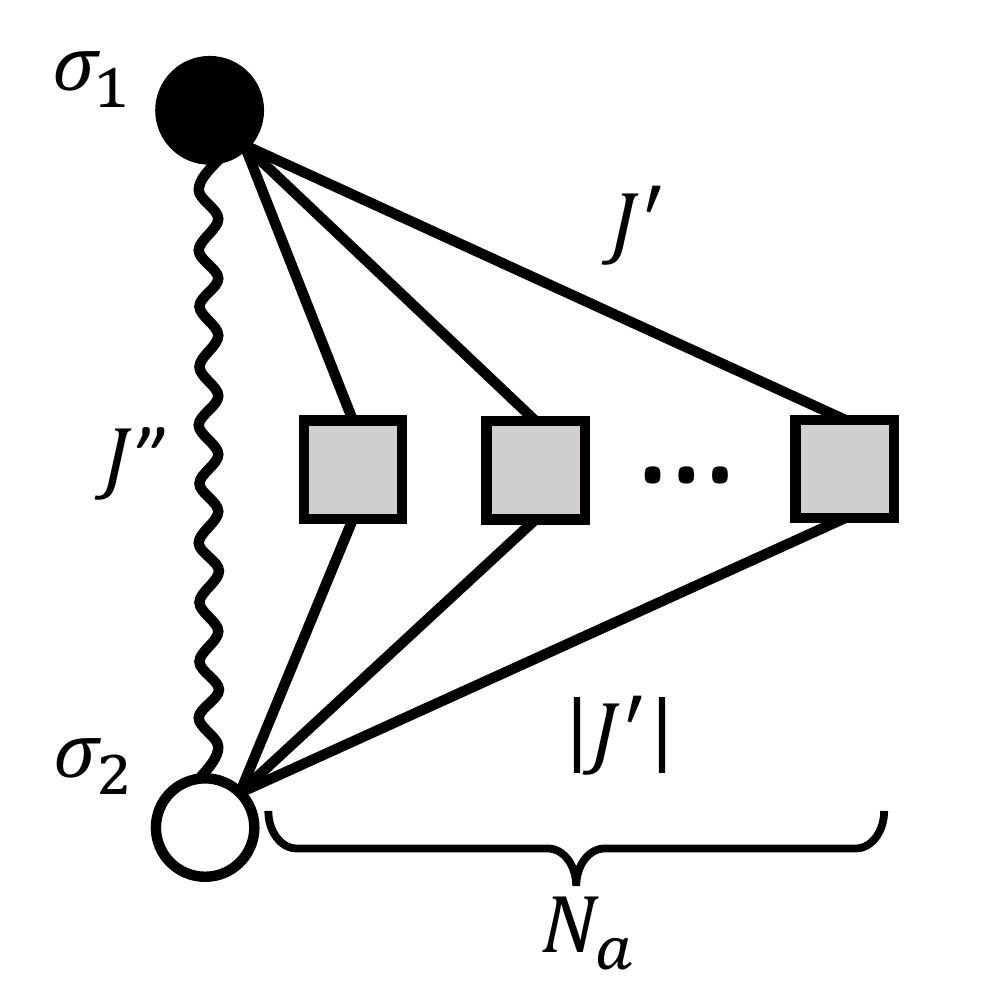}
    \label{fig:posi_antipara}
  }
  \caption{
  (a) Parallel and (b) antiparallel system spin cases where the original interactions are positive. Black, white, and gray symbols denote $+$, $-$, and disordered spins, respectively. Circles and squares denote the system and auxiliary spins, respectively.}
  \label{fig:posi_para_antipara}
\end{figure}

First, we analyzed the case where the original interaction is positive (Fig.~\ref{fig:posi_para_antipara}). 
Figs.~\ref{fig:posi_para} and~\ref{fig:posi_antipara} depict ``the parallel state'' (e.g., $(\sigma_1, \sigma_2) = (+, +)$) and ``antiparallel state'' of system spins (e.g., $(\sigma_1, \sigma_2) = (+, -)$), respectively.
Let $m$ denote the number of auxiliary spins that represent the $+$ internal field when the original interaction is positive. 
The energies of the parallel state and antiparallel states are given by
\begin{align}
  E^{(N_\mathrm{a})}_{++}(m)=(2N_\mathrm{a}-4m)J'-J'',
  \label{eq:posi_para_energy}
\end{align}
\begin{align}
  E^{(N_\mathrm{a})}_{+-}(m)=J''.
  \label{eq:posi_antipara_energy}
\end{align}

We considered the probability distribution of the auxiliary spins at a temperature $T$.
In the parallel and antiparallel states, each probability of $m$ up spins in the $N_\textrm{a}$ auxiliary spins are given by
\begin{align}
  Q^{(N_\mathrm{a})}_{++}(m)=\frac{\exp(-2\beta{J'}N_\mathrm{a})}{(2\cosh{2\beta{J'}})^{N_\mathrm{a}}}\binom{N_\mathrm{a}}{m}\exp(4\beta{J'}m),
  \label{eq:posi_para_probability}
\end{align}
\begin{align}
  Q^{(N_\mathrm{a})}_{+-}(m)=\binom{N_\mathrm{a}}{m}\left(\frac{1}{2}\right)^{N_\mathrm{a}}.
  \label{eq:posi_antipara_probability}
\end{align}

\begin{figure}[b]
  \centering
  \subfigure[]{
    \includegraphics[clip,width=0.4\linewidth]{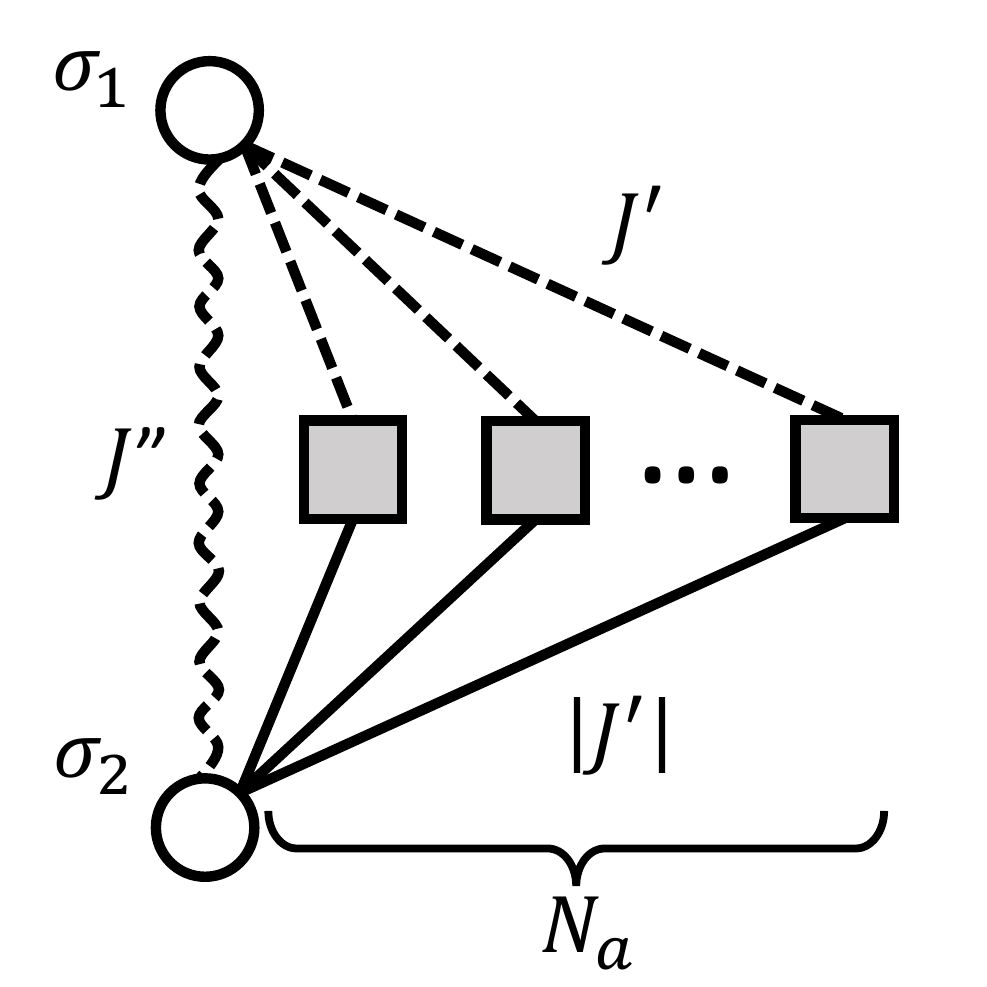}
    \label{fig:nega_para}
  }
  \subfigure[]{
    \includegraphics[clip,width=0.4\linewidth]{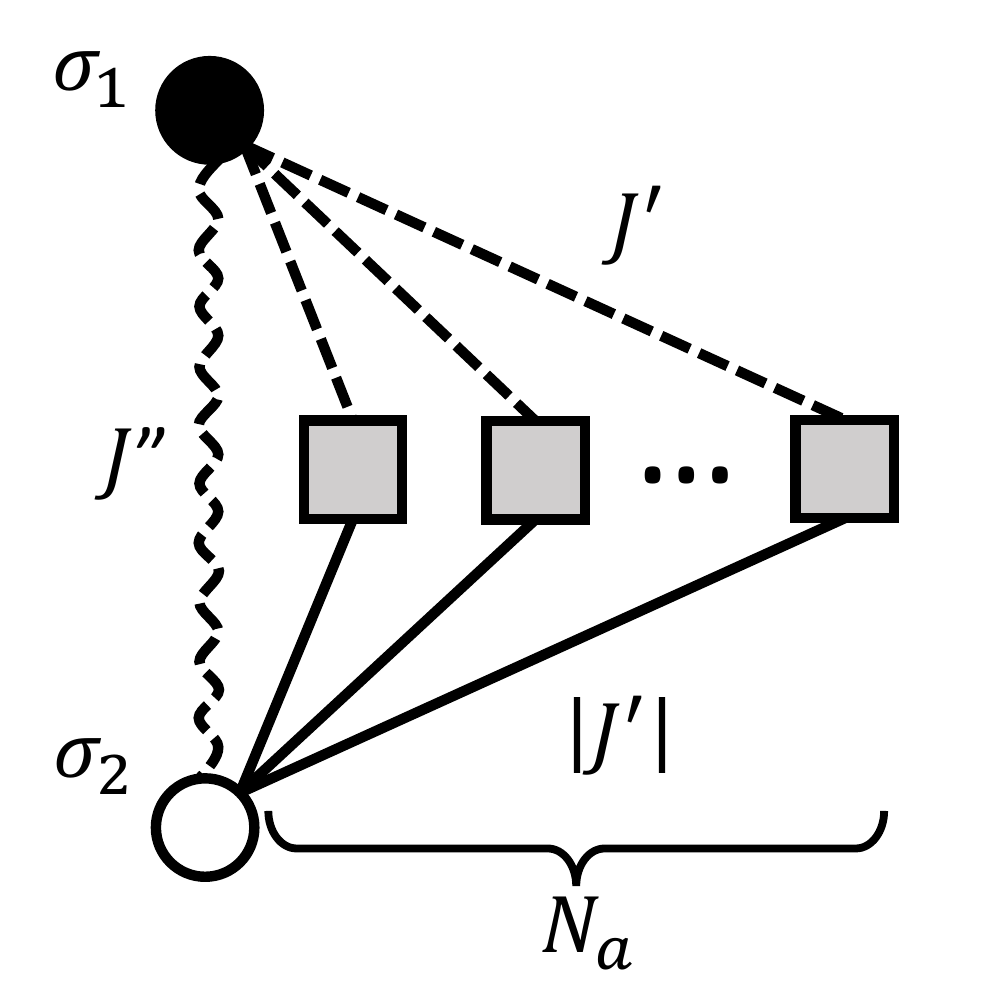}
    \label{fig:nega_antipara}
  }
  \caption{
  (a) Parallel and (b) antiparallel system spin cases where the original interactions are negative. Black, white, and gray symbols denote $+$, $-$, and disordered spins, while the solid and dotted lines denote positive and negative, respectively. Circles and squares denote the system spins and the auxiliary spins, respectively.}
  \label{fig:nega_para_antipara}
\end{figure}

Next, we analyzed the case where the original interaction is negative.
Figs.~\ref{fig:nega_para} and ~\ref{fig:nega_antipara} depict ``the parallel state'' of system spins (e.g., $(\sigma_1, \sigma_2) = (-, -)$) and ``antiparallel state'' of system spins (e.g., $(\sigma_1, \sigma_2) = (+, -)$), respectively. 
Let $n$ be the number of auxiliary spins that represent the $+$ internal field when the original interaction is negative.
The energies of the parallel and antiparallel states are given by 

\begin{align}
  E^{(N_\mathrm{a})}_{--}(n)=-J'',
  \label{eq:nega_para_energy}
\end{align}
\begin{align}
  E^{(N_\mathrm{a})}_{+-}(n)=(2N_\mathrm{a}-4n)J'+J''.
  \label{eq:nega_antipara_energy}
\end{align}

Each probability of $n$ up spins in the $N_\textrm{a}$ auxiliary spins of the parallel and antiparallel state are given by
\begin{align}
  R^{(N_\mathrm{a})}_{--}(n)=\binom{N_\mathrm{a}}{n}\left(\frac{1}{2}\right)^{N_\mathrm{a}},
  \label{eq:nega_para_probability}
\end{align}
\begin{align}
  R^{(N_\mathrm{a})}_{+-}(n)=\frac{\exp(-2\beta{J'}N_\mathrm{a})}{(2\cosh{2\beta{J'}})^{N_\mathrm{a}}}\binom{N_\mathrm{a}}{n}\exp(4\beta{J'}n).
  \label{eq:nega_antipara_probability}
\end{align}

Since $Q^{(N_\textrm{a})}_{+-}(m)$ and $R^{(N_\textrm{a})}_{--}(n)$ are independent of the temperature, the equations are simple binomial distributions and equivalent. 
In contrast, $Q^{(N_\textrm{a})}_{++}(m)$ and $R^{(N_\textrm{a})}_{+-}(n)$ depend on temperature. 
They are maximized at $N_\textrm{a}/2$ for high temperatures due to the entropy effect, whereas they are maximized at nearly $N_\textrm{a}$ and $0$ for low temperatures. 

\begin{figure}[t]
  \centering
  \subfigure[]{
    \includegraphics[clip,width=0.45\linewidth]{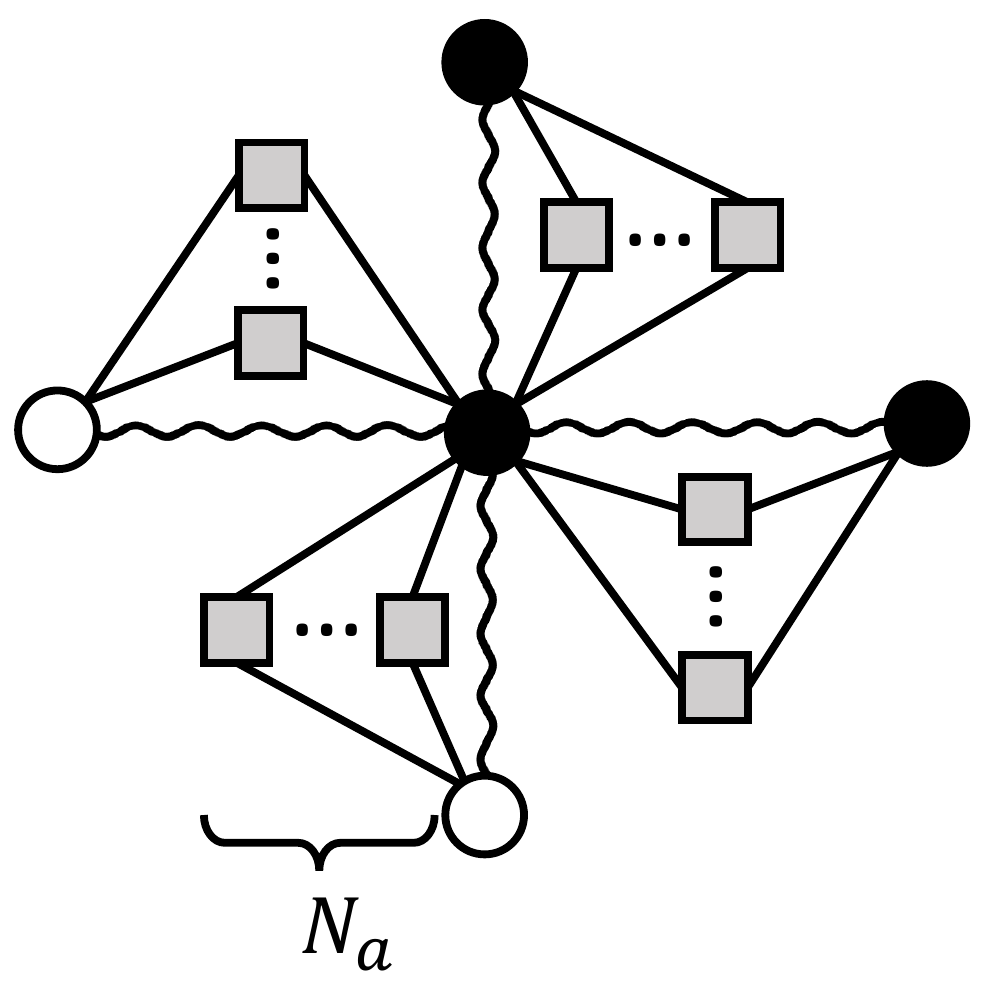}
    \label{fig:posi_free_spin}
  }
  \subfigure[]{
    \includegraphics[clip,width=0.45\linewidth]{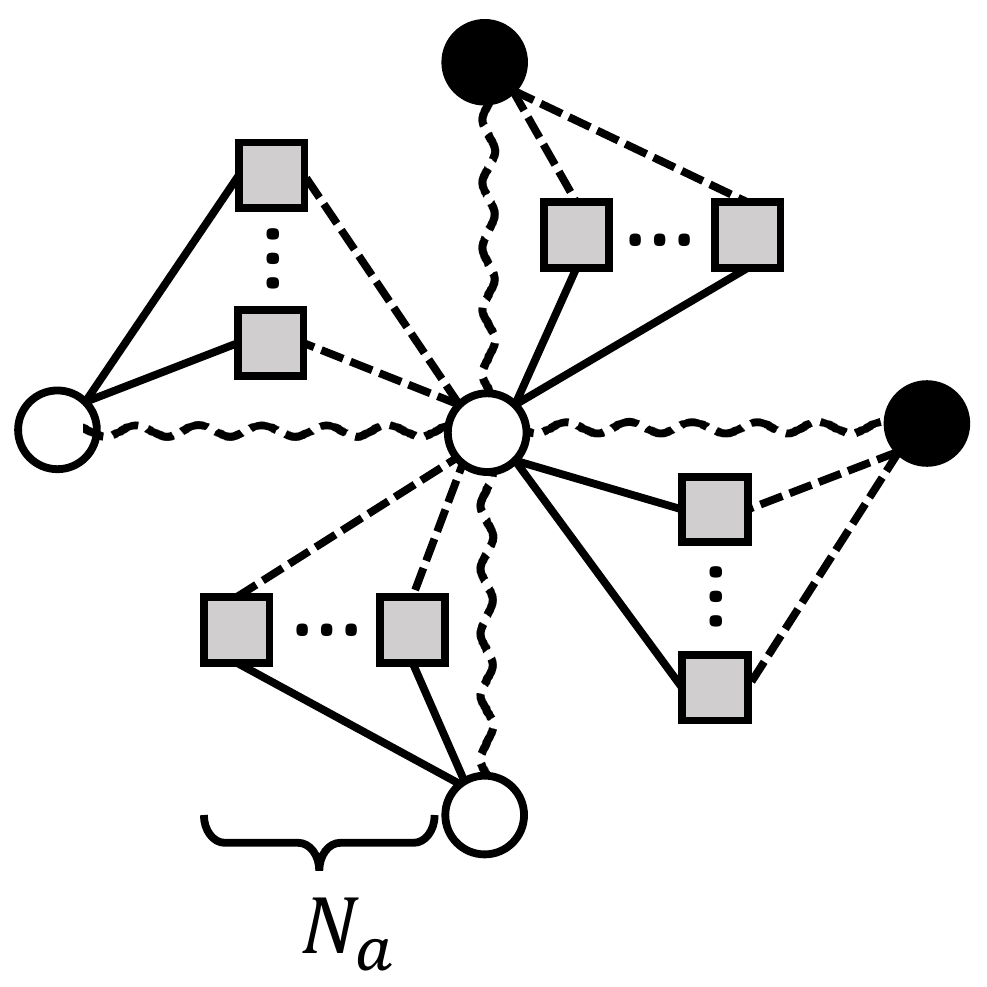}
    \label{fig:nega_free_spin}
  }
  \caption{Model for analyzing the flip probability of the central spin (free spin) where the original interactions are (a) positive and (b) negative. Black, white, and gray symbols denote $+$, $-$, and disordered spins, respectively. Circles and squares denote the system and auxiliary spins, respectively.}
  \label{fig:free_spin}
\end{figure}

To estimate the transition probability of the system spins on the BWR Ising model, we calculated the flip probability of the central spin shown in Fig.~\ref{fig:free_spin} following~\cite{Tanaka2009}. 
The central spin, which we refer to as ``free spin,'' is surrounded by two up spins and two down spins.
When $N_\textrm{a}=0$, the flip probability of the free spin is $1/2$ in the Glauber dynamics~\cite{glauber1963time}.
However, when auxiliary spins are added by the proposed method ($N_\textrm{a}>0$), the flip probability becomes less than $1/2$ due to the distribution of the surrounding auxiliary spins. 

Fig.~\ref{fig:posi_free_spin} shows the case where the original interaction is positive. 
The internal field on the free spin is given by 
\begin{align}
  h(n_1, n_2)=2J'(2N_\mathrm{a}-n_1-n_2),
  \label{eq:posi_inter}
\end{align}
where $n_1$ and $n_2$ are the numbers of auxiliary spins representing the $+$ internal field in the parallel and antiparallel state, respectively.
In the Glauber dynamics, the flip probability of free spin $P_\textrm{flip}$ is given by 
\begin{multline}
  P_\mathrm{flip}=\sum_{(n_1, n_2)}Q^{(2N_\mathrm{a})}_{++}(n_1)Q^{(2N_\mathrm{a})}_{+-}(n_2)\\
  \times\frac{1}{1+\exp[-2\beta{h}(n_1, n_2)]},
  \label{eq:posi_pflip_1}
\end{multline}
\begin{multline}
  P_\mathrm{flip}=\frac{\exp(-4\beta{J'}N_\mathrm{a})}{(4\cosh{2\beta{J'}})^{2N_\mathrm{a}}}\sum_{(n_1, n_2)}\binom{2N_\mathrm{a}}{n_1}\binom{2N_\mathrm{a}}{n_2}\\
  \times\frac{\exp(4\beta{J'}n_1)}{1+\exp[-4\beta{J'}(2N_\mathrm{a}-n_1-n_2)]}.
  \label{eq:posi_pflip_2}
\end{multline}

Similarly, in the case where the original interaction is negative (Fig.~\ref{fig:nega_free_spin}), the internal field on a free spin is given by
\begin{align}
  h(n_3, n_4)=-2J'(n_3-n_4),
  \label{eq:nega_inter}
\end{align}
where $n_3$ and $n_4$ are the numbers of auxiliary spins representing the $+$ internal field in the antiparallel and parallel states, respectively.
In the Glauber dynamics, the $P_\textrm{flip}$ is given by 
\begin{multline}
  P_\mathrm{flip}=\sum_{(n_3, n_4)}R^{(2N_\mathrm{a})}_{+-}(n_3)R^{(2N_\mathrm{a})}_{--}(n_4)\\
  \times\frac{1}{1+\exp[-2\beta{h}(n_3, n_4)]},
  \label{eq:nega_pflip_1}
\end{multline}
\begin{multline}
  P_\mathrm{flip}=\frac{\exp(-4\beta{J'}N_\mathrm{a})}{(4\cosh{2\beta{J'}})^{2N_\mathrm{a}}}\sum_{(n_3, n_4)}\binom{2N_\mathrm{a}}{n_3}\binom{2N_\mathrm{a}}{n_4}\\
  \times\frac{\exp(4\beta{J'}n_3)}{1+\exp[4\beta{J'}(n_3-n_4)]}.
  \label{eq:nega_pflip_2}
\end{multline}

The $P_\textrm{flip}$ is the same for arbitrary system spin combinations as in (\ref{eq:posi_pflip_2}) and (\ref{eq:nega_pflip_2}) .

\begin{figure}[t]
  \centering
  \includegraphics[clip,width=0.9\linewidth]{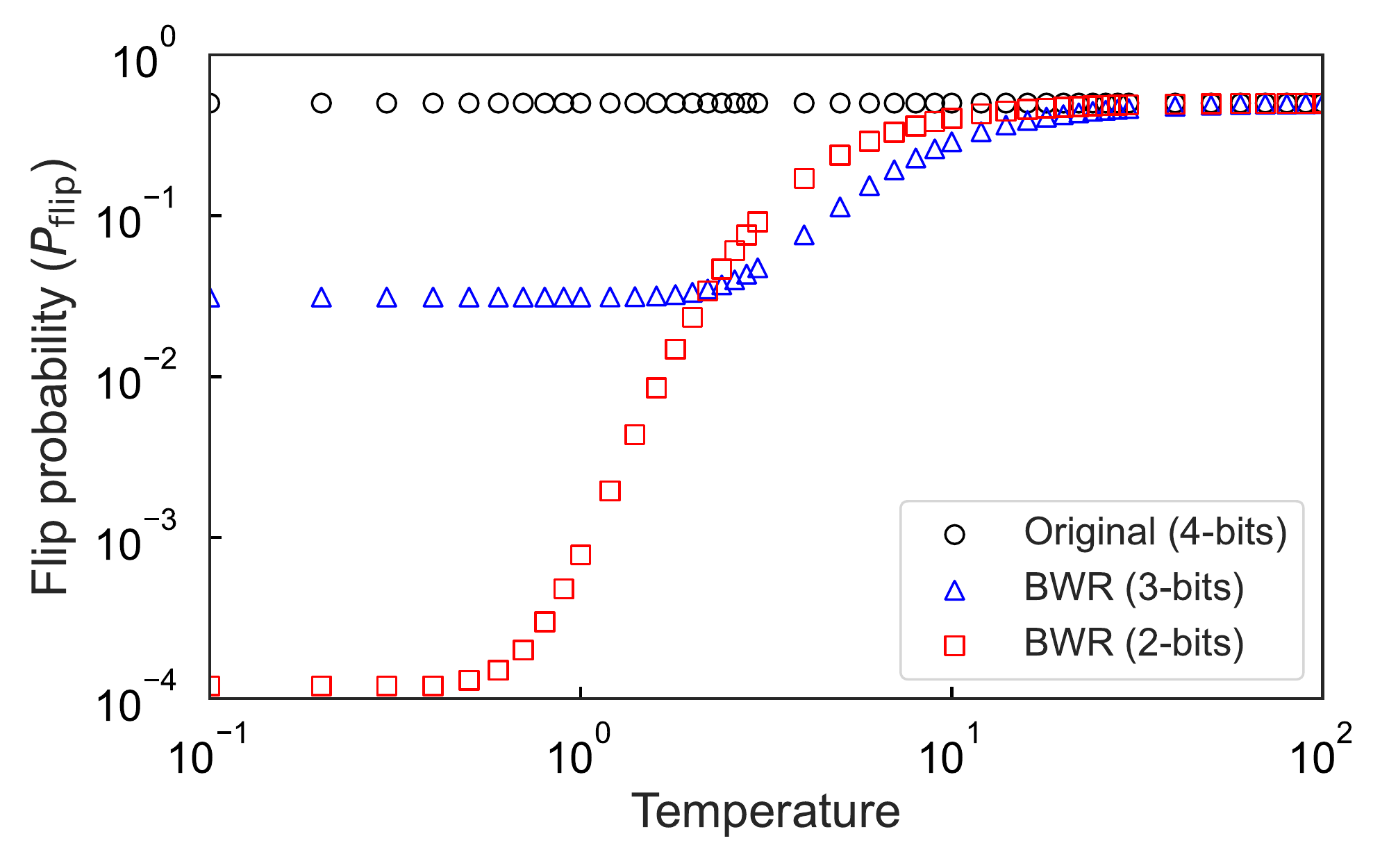}
  \caption{Flip probability of the free spin of the original Ising model ($J=7$) and the BWR Ising model for several temperatures. Red squares, blue triangles, and black circles denote the BWR Ising model ($2$-bits), BWR Ising model ($3$-bits), and original Ising model ($4$-bits), respectively.}
  \label{fig:pflip_comp}
\end{figure}

Fig.~\ref{fig:pflip_comp} compares the flip probability of the free spin with the original and BWR Ising models in the previous section.
Although the probability is constant with the number of auxiliary spins $N_\textrm{a}$ at high temperatures, it changes significantly at low temperatures.
Note that there is a limit to $P_\textrm{flip}$ at low temperatures because the slow relaxation is caused by the entropy effect. 
The value of the limit can be expressed as  
\begin{align}
  \lim_{T \to 0}P_\mathrm{flip}=\frac{1}{2}\left(\frac{1}{4}\right)^{N_\mathrm{a}}.
  \label{eq:pflip_0}
\end{align}

A discrepancy in the flip probability occurs between the original and BWR Ising models at low temperatures, which was not considered in the parameters for SA in the previous section.
This discrepancy likely affects the difference in the dynamical processes. 
Note that the entropy effect does not occur in the auxiliary spins for the magnetic fields.  

\section{Proposed SA parameters}
\label{sec:proposed_parameter}

In the previous section, we analyzed the BWR Ising model using the proposed method. 
The BWR Ising model has two characteristic properties: an effective temperature and a slow relaxation. 
These properties are not present in the original Ising model. 
In Section~\ref{sec:simulation}, it was speculated that the dynamical processes between the original and the BWR Ising model differ because the SA is performed with the same SA parameters before and after bit-width reduction, despite the variation in the statistical mechanics properties. 
This section proposes SA parameters that consider the properties of the BWR Ising model and evaluate the proposed SA parameters experimentally.

\subsection{How to modify the parameters}
\label{subsec:how_to}

\begin{figure*}[t]
  \centering
  \subfigure[]{
    \includegraphics[clip,width=0.4\linewidth]{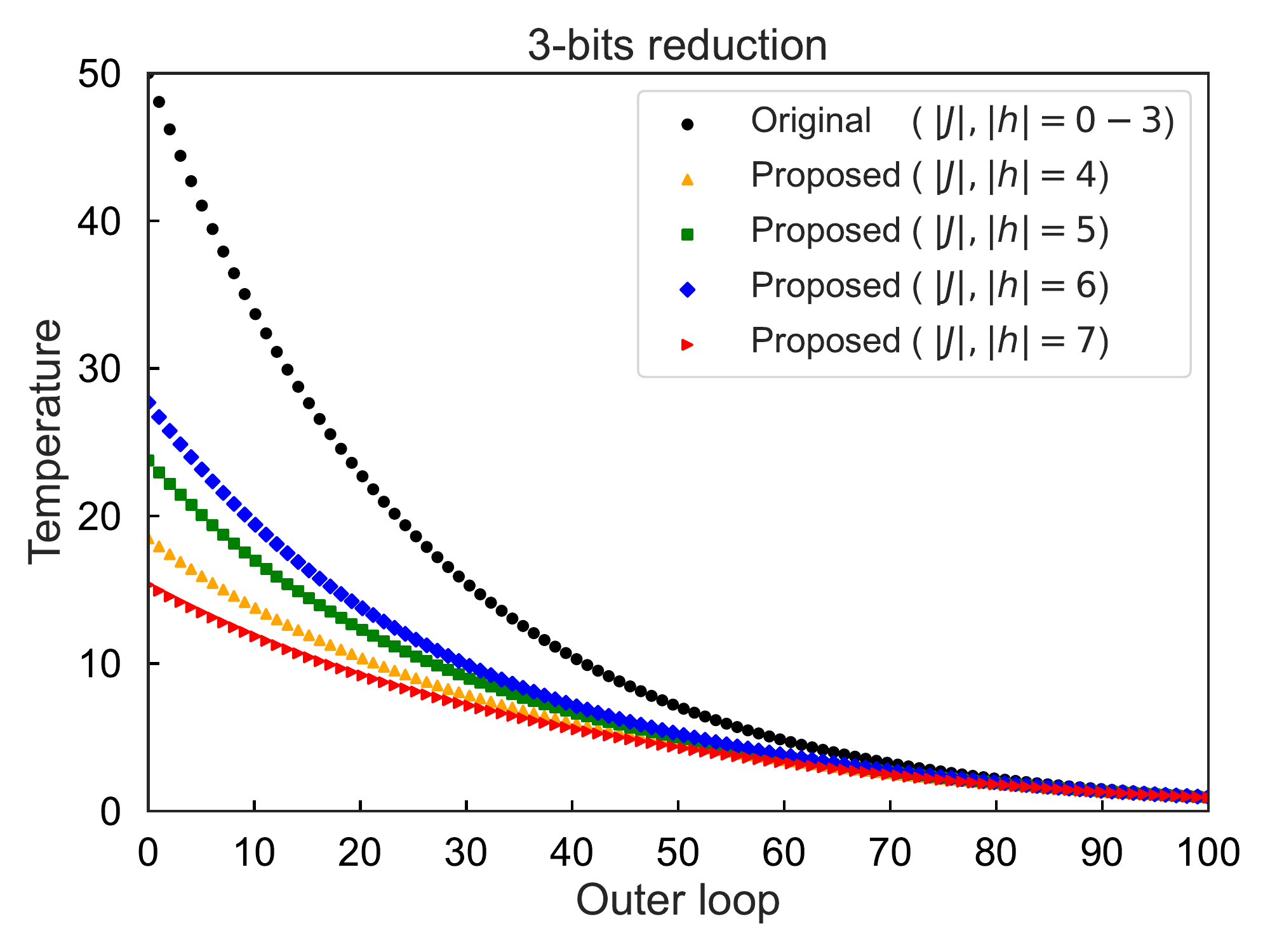}
    \label{fig:proposed_temp_3bits}
  }
  \subfigure[]{
    \includegraphics[clip,width=0.4\linewidth]{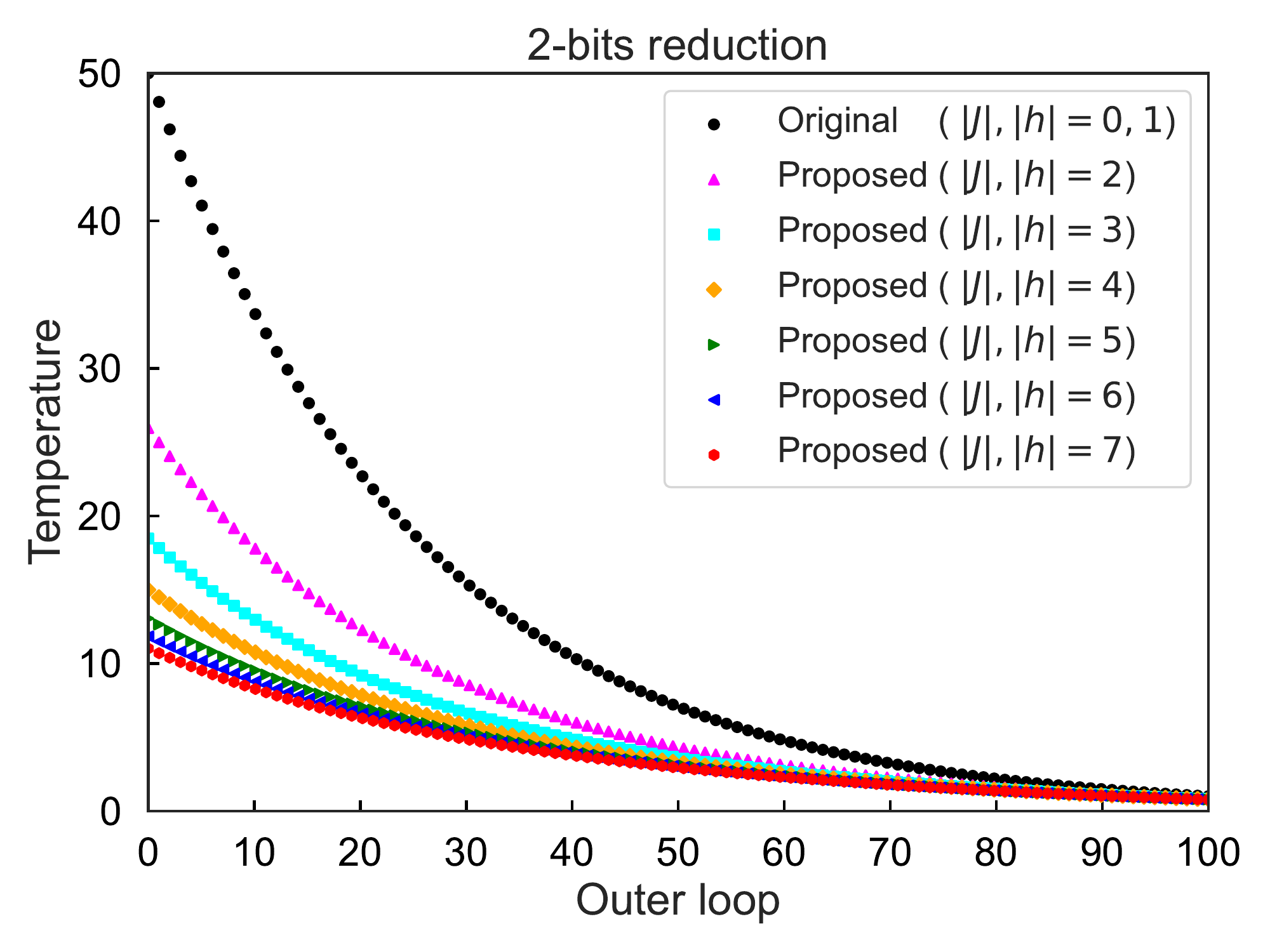}
    \label{fig:proposed_temp_2bits}
  }
  \caption{
    Proposed temperature schedules of the BWR Ising model for several coefficients of the magnetic fields or interactions. Bit-widths are reduced to (a) $3$-bits and (b) $2$-bits. 
    }
  \label{fig:proposed_temp}
\end{figure*}

First, the temperature schedule is modified based on the effective temperature $T_\textrm{eff}$ so that $T_\textrm{eff}$ is closer to the temperature $T$ of the original temperature schedule using~(\ref{eq:mag_posi_Teff}),~(\ref{eq:mag_nega_Teff}),~(\ref{eq:int_posi_Teff}), or~(\ref{eq:int_nega_Teff}). 
Fig.~\ref{fig:proposed_temp} shows the original temperature schedule used in Section~\ref{sec:simulation} and the proposed temperature schedules when the absolute value of the coefficient 0-7 is reduced to $3$- or $2$-bits.

Next, the inner loop is modified based on the flip probability. 
To realize a flip probability of the BWR Ising model closer to that of the original Ising model, we define an effective relaxation time $\tau_\textrm{eff}$.
According to a previous study~\cite{Tanaka2009}, $\tau_\textrm{eff}$ is given by
\begin{align}
  \tau_\mathrm{eff}={P_\mathrm{flip}}^{-1}.
  \label{eq:effective_time}
\end{align}

\begin{figure*}[t]
  \centering
  \subfigure[]{
    \includegraphics[clip,width=0.4\linewidth]{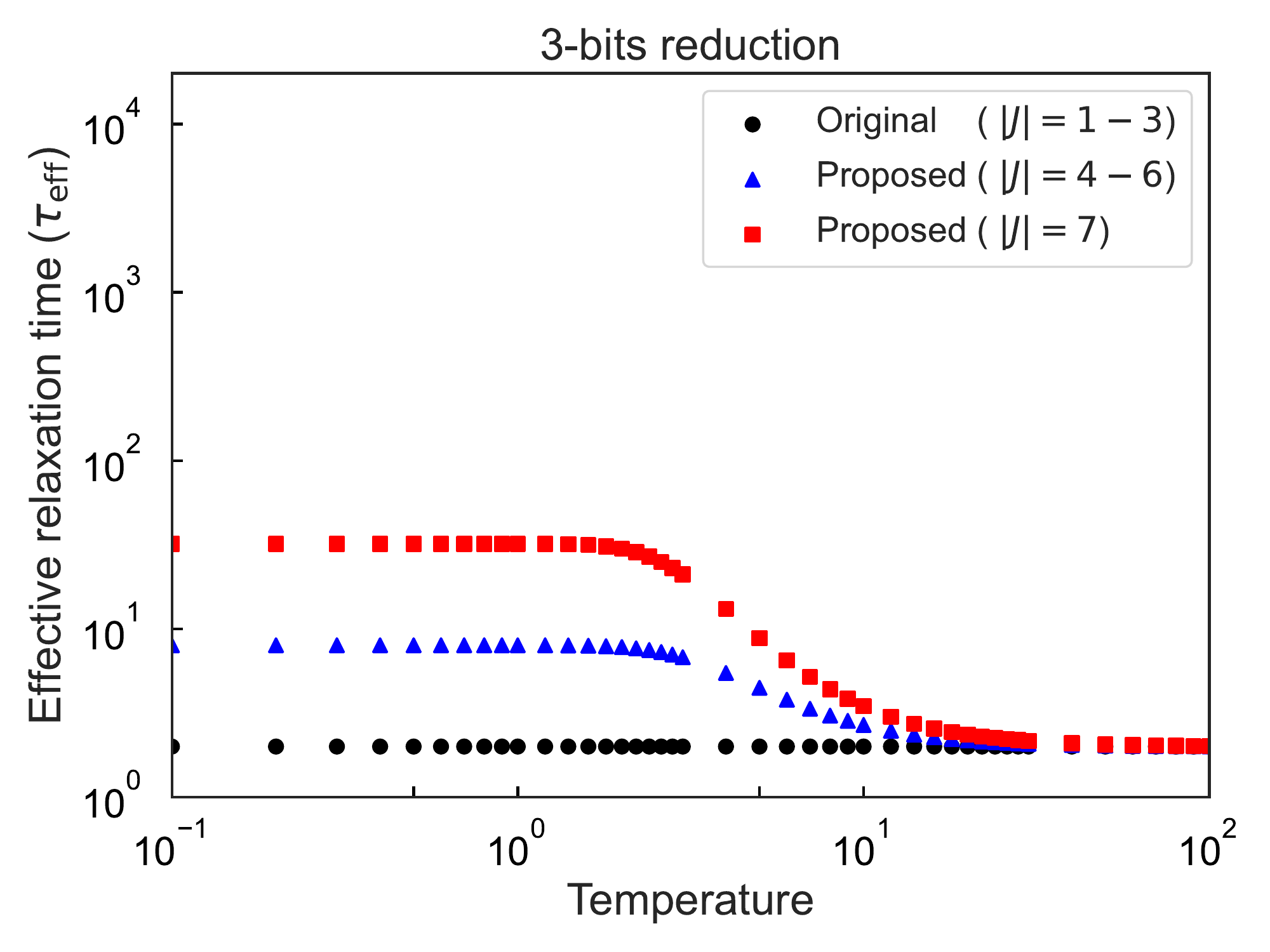}
    \label{fig:proposed_inner_3bits}
  }
  \subfigure[]{
    \includegraphics[clip,width=0.4\linewidth]{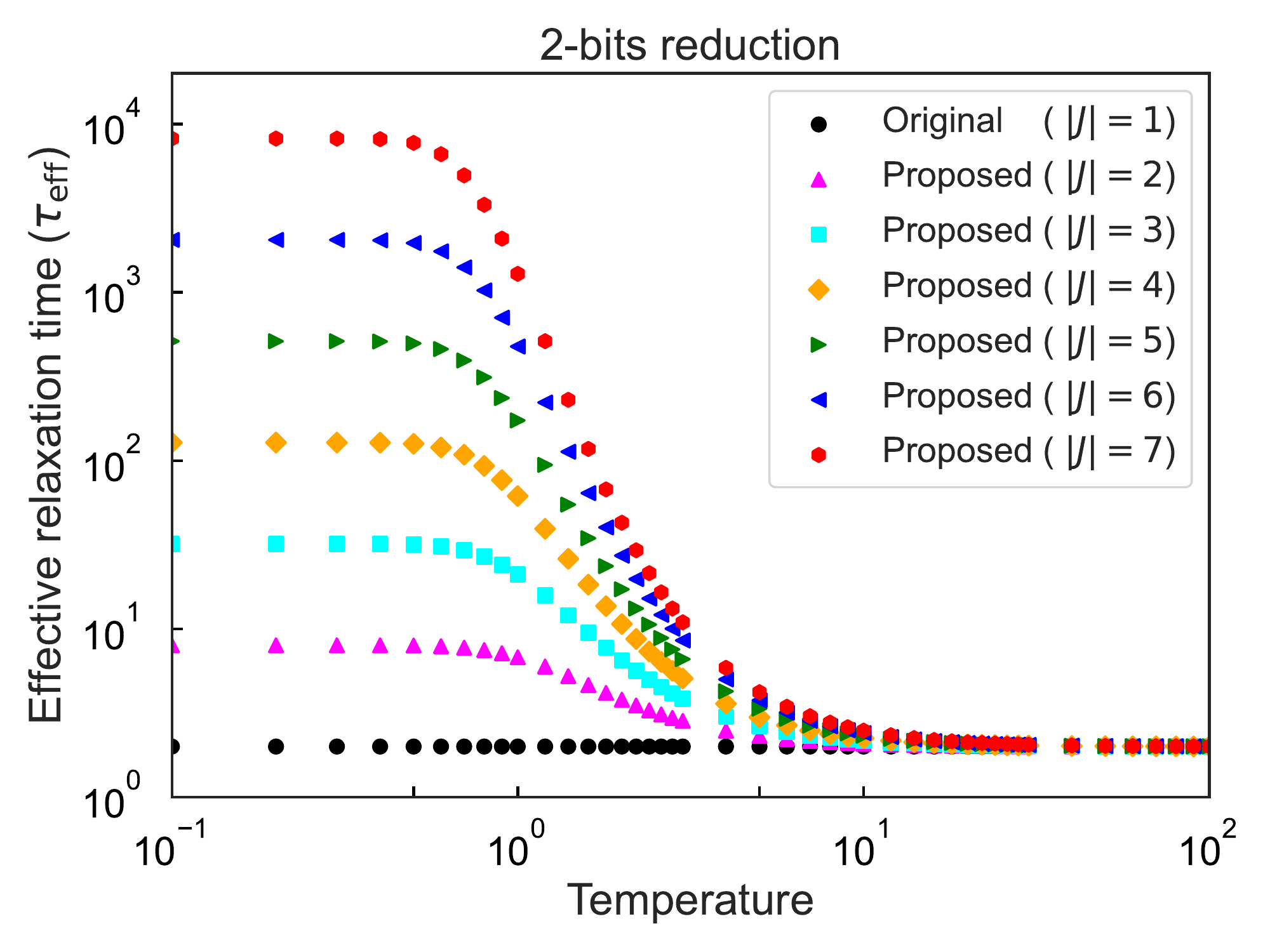}
    \label{fig:proposed_inner_2bits}
  }
  \caption{
    Effective time relaxation $\tau_\textrm{eff}$ of the BWR Ising model for several coefficients of interactions. Bit-widths are reduced to (a) $3$-bits width and (b) $2$-bits width.
  }
  \label{fig:proposed_inner}
\end{figure*}

Fig.~\ref{fig:proposed_inner} shows the relationship between temperature $T$ and $\tau_\textrm{eff}$ of the original or the BWR Ising model.
$\tau_\textrm{eff}$ of the original Ising model is two from the definition of the system shown in Fig.~\ref{fig:free_spin}. 
The absolute value of coefficient 0-7 is reduced to $3$- or $2$-bits.
Then $\tau_\textrm{eff}$ can be calculated by~(\ref{eq:posi_pflip_2}) or~(\ref{eq:nega_pflip_2}), and~(\ref{eq:effective_time}).

The algorithm to modify the parameters is as follows:

\begin{description}
   \item[Step 1:]Set the temperature schedule so that $T_\textrm{eff}$ is closer to the original one.
   \item[Step 2:]Set the inner loop so that $\tau_\textrm{eff}$ of the BWR Ising model is closer to that of the original Ising model. $\tau_\textrm{eff}$ is obtained at each temperature determined in step 1. The proposed inner loop is set $1$ MCS$\times \tau_\textrm{eff}/2$ because the original inner loop corresponds to $\tau_\textrm{eff}=2$.
\end{description}

\subsection{experimental evaluation}
\label{subsec:experimental_evaluation}
\begin{figure}[t]
  \centering
  \includegraphics[clip,width=0.9\linewidth]{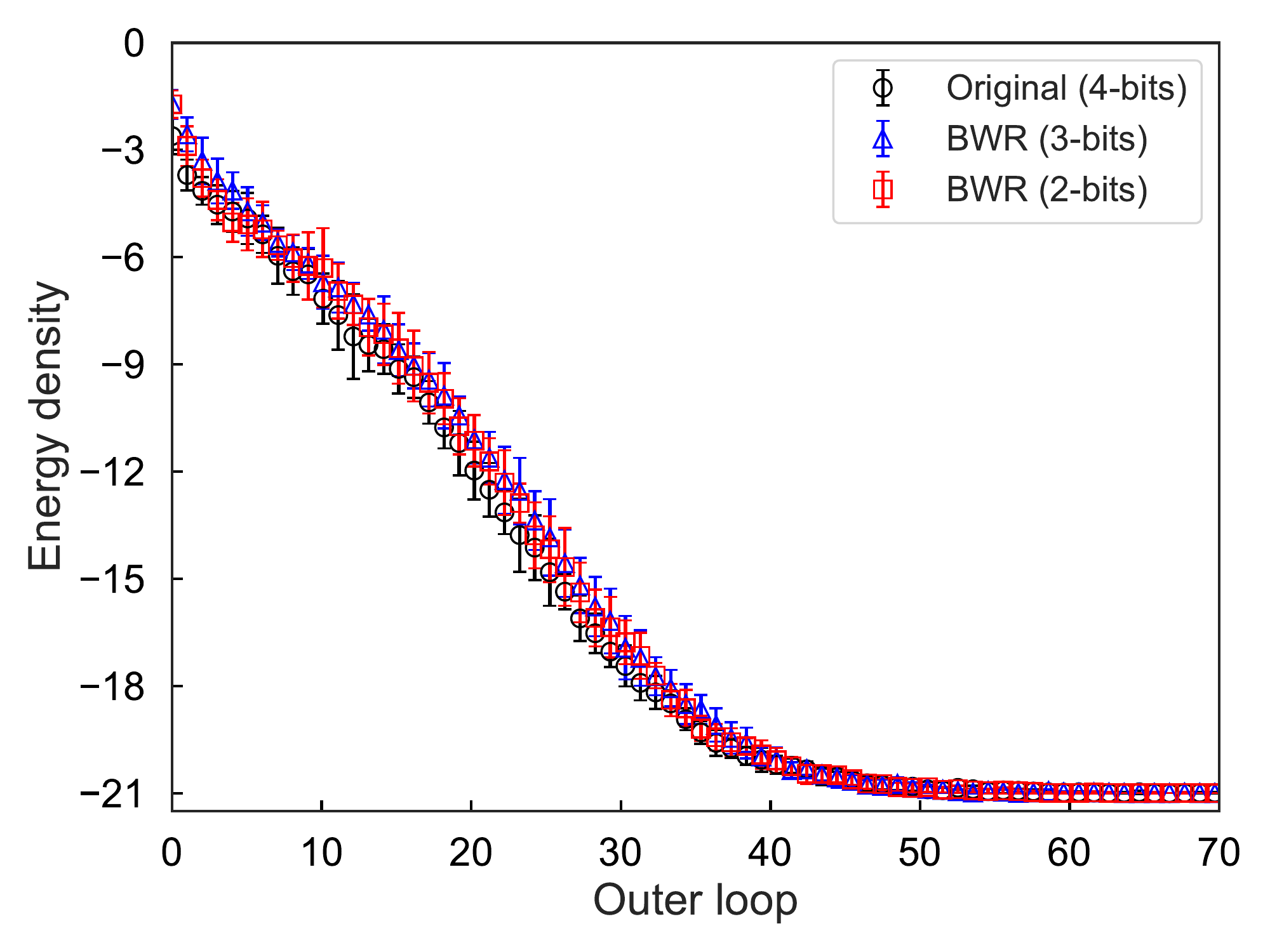}
  \caption{
  Dynamical process of the BWR Ising model with the proposed SA parameters and the original Ising model with the original SA parameters. Red squares, blue triangles, and black circles denote the BWR Ising model ($2$-bits), BWR Ising model ($3$-bits), and the original Ising model ($4$-bits), respectively. Every plot is the average of ten runs. The error bars are standard deviations.}
  \label{fig:simulation_proposed}
\end{figure}

To evaluate the effectiveness of the proposed SA parameters, we experimentally investigated the dynamical process of the energy density on the Ising model used in Section~\ref{sec:simulation}.
We performed SA of the original Ising model with the SA parameters described in Table~\ref{table:parameters_simulation}. 
For the BWR Ising model, the SA parameters were changed from Table~\ref{table:parameters_simulation} to the proposed temperature schedule and the inner loop explained in this section.
We call this the ``proposed SA parameters.''
The coefficients were reduced to $3$- or $2$-bits. 
Fig.~\ref{fig:simulation_proposed} shows the results.
The dynamical process of the BWR Ising model with the proposed SA parameters is similar to that of the original Ising model.
Similar effects were observed in the large-size square lattice systems ($L=40, 50$) in Appendix~\ref{sec:appendixB}.

\section{numerical results}
\label{sec:random_ising}

We evaluated the applicability of the proposed SA parameters when the Ising model has random coefficients.
We compared the dynamical process of the original and BWR Ising models using random Ising models~\cite{instances}. 
We performed SA of a square lattice system, where $L=30$.
The coefficients of the magnetic fields and interactions take integer values from $[-7, 7]$ with equal probabilities. 
Although $0$ was excluded for the interactions, it was included for the magnetic fields. 
In these demonstrations, the bit-width of the original Ising model was reduced from $4$-bits to $3$- or $2$-bits. 

Table~\ref{table:parameters_simulation} shows the SA parameters in this demonstration, except for the cooling rate.
The cooling rate $r$ was set such that the final temperature was equal to $1$, i.e. $r = 0.9612$.
This condition ensures that the final temperature is sufficiently small relative to the coefficients of the original Ising model. 
See Appendix~\ref{sec:appendixB} for results using different types of temperature schedules.

In this demonstration, we performed SA with four types of SA parameters. 

\begin{itemize}
  \item \textit{Original SA parameter}\\
  The unmodified SA parameters. 
  \item \textit{Modified temperature schedule (TS) SA parameter}\\
  Only the temperature schedule is modified based on the maximum absolute value of the coefficients (i.e., $|J|, |h|=7$). A modified temperature schedule based on the maximum absolute value shows the most gradual temperature decrease from the lowest temperature (Fig.~\ref{fig:proposed_temp}). 
  \item \textit{Modified inner loop (IL) SA parameter}\\
  Only the inner loop is modified based on the maximum absolute value of the coefficients. Effective relaxation time $\tau_\textrm{eff}$ based on the maximum absolute value shows the longest effective relaxation time (Fig.~\ref{fig:proposed_inner}).
  \item \textit{Proposed SA parameter}\\
  Both the temperature schedule and the inner loop are modified. 
\end{itemize}

\begin{figure*}[t]
  \centering
  \subfigure[]{
    \includegraphics[clip,width=0.45\linewidth]{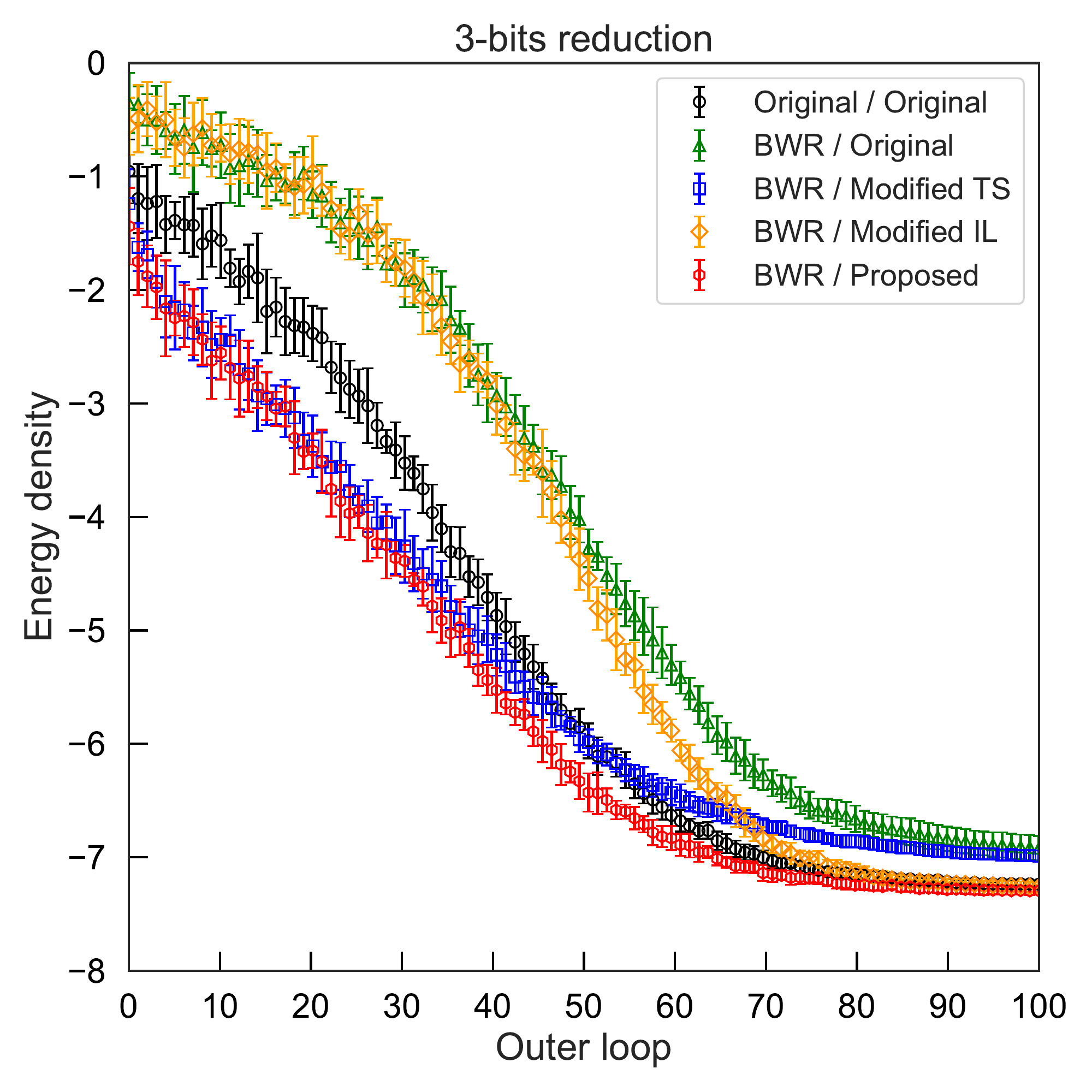}
    \label{fig:dynamics_random_3bits}
  }
  \subfigure[]{
    \includegraphics[clip,width=0.45\linewidth]{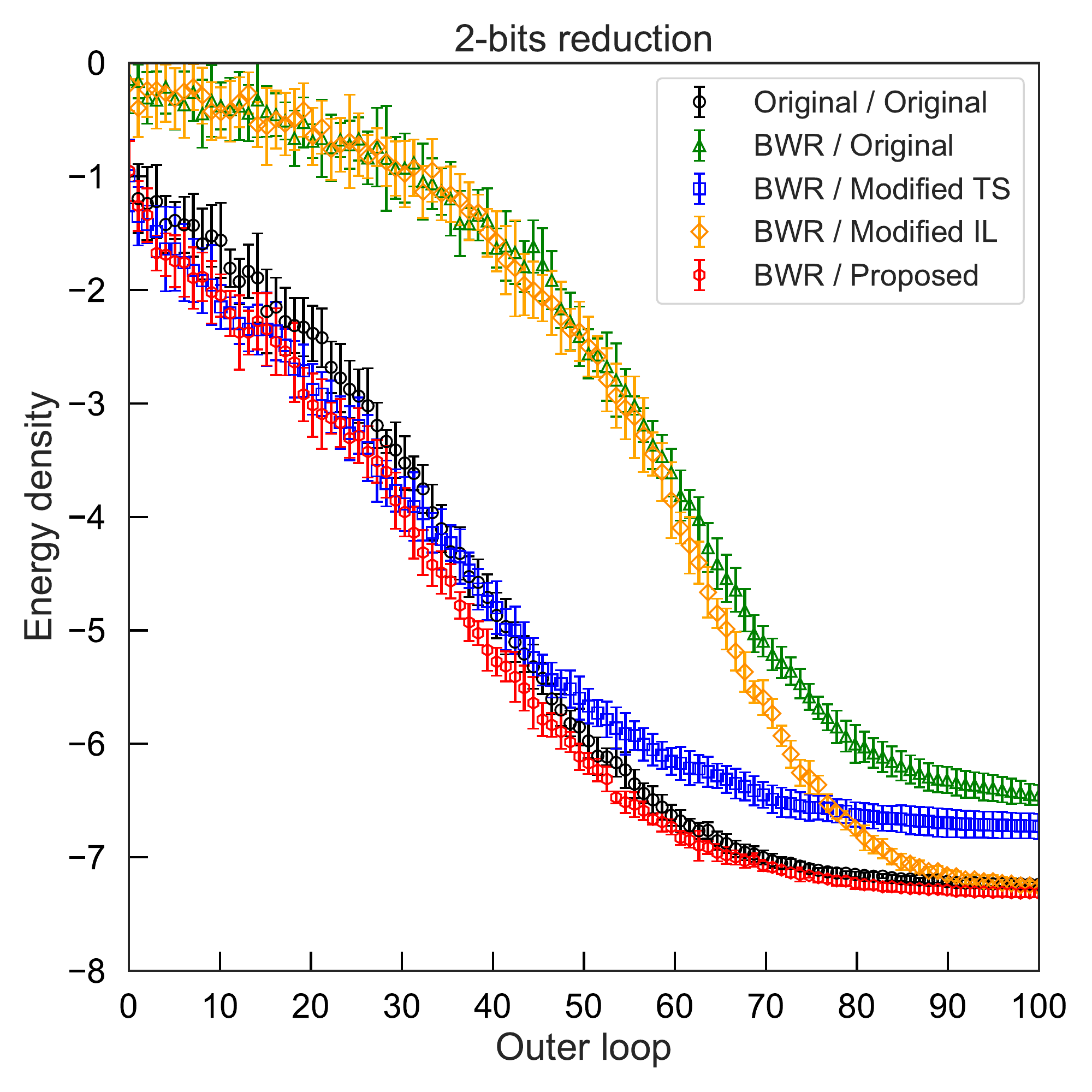}
    \label{fig:dynamics_random_2bits}
  }
  \caption{
    Dynamical processes of the BWR Ising model and the original Ising model. Black circles denote the original Ising model. Green triangles, blue squares, orange diamonds, and red hexagons denote the BWR Ising model with the original SA parameters, modified temperature schedule only, modified inner loop only, and proposed SA parameters, respectively. Every plot is the average of ten runs. The error bars are standard deviations.
  }
  \label{fig:dynamics_random}
\end{figure*}

Fig.~\ref{fig:dynamics_random} shows the dynamical processes of the original and BWR Ising models for each set of SA parameters.
The data represent the average and standard deviation of the energy density for ten SA simulations.
By modifying the temperature schedule, the energy density of the BWR Ising model becomes closer to that of the original Ising model at the beginning of the iteration.
Applying an effective relaxation time to the inner loop prevents the energy density of the BWR Ising model from terminating at a high value in the later stages of the iteration.
The dynamical processes of the BWR Ising model with the proposed SA parameters are almost the same as that of the original Ising model.
However, a difference appears in the early stages of the iterations when reducing to $3$-bits.

\begin{figure*}[t]
  \centering
  \subfigure[]{
    \includegraphics[clip,width=0.23\linewidth]{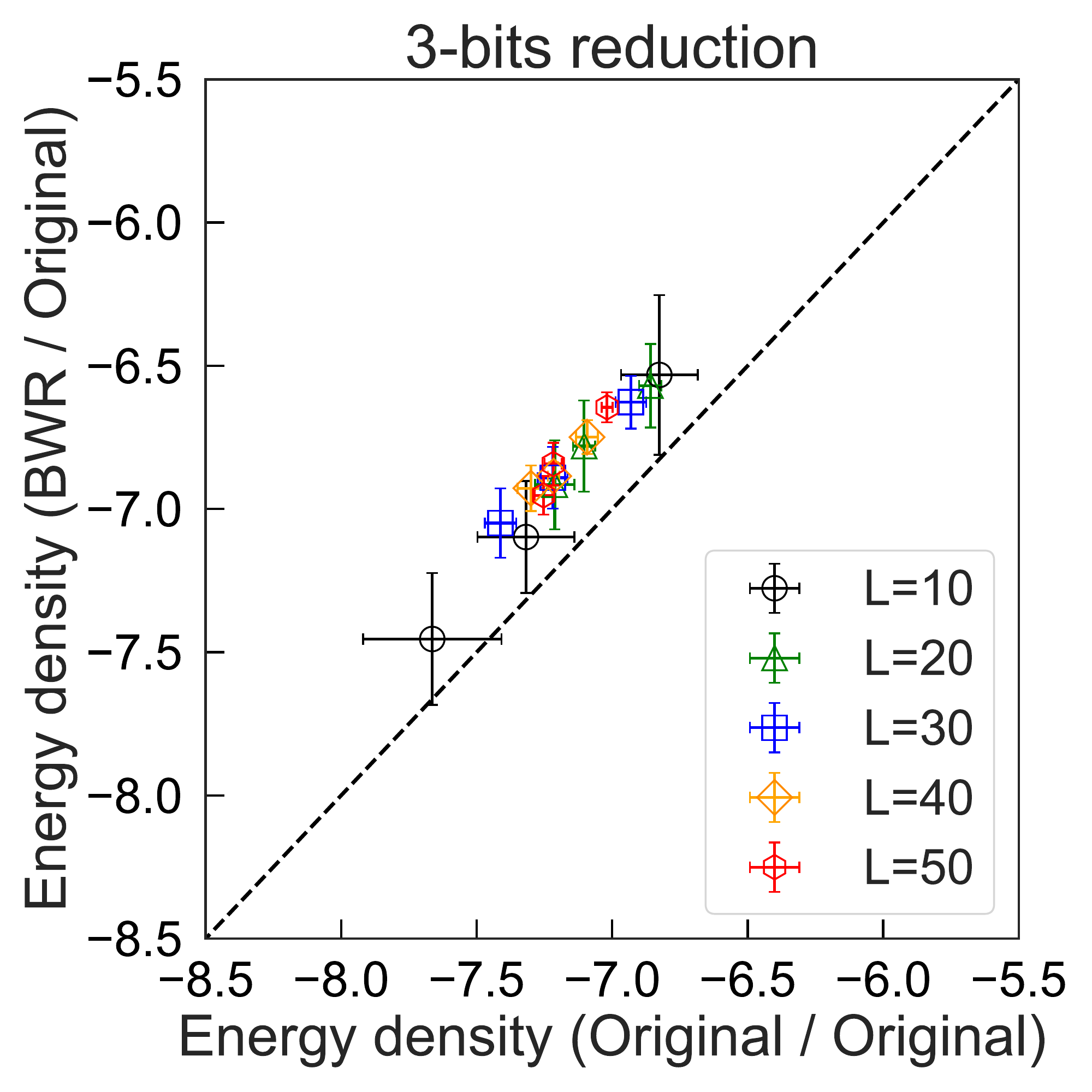}
    \label{fig:scatter_3bits_original}
  }
  \subfigure[]{
    \includegraphics[clip,width=0.23\linewidth]{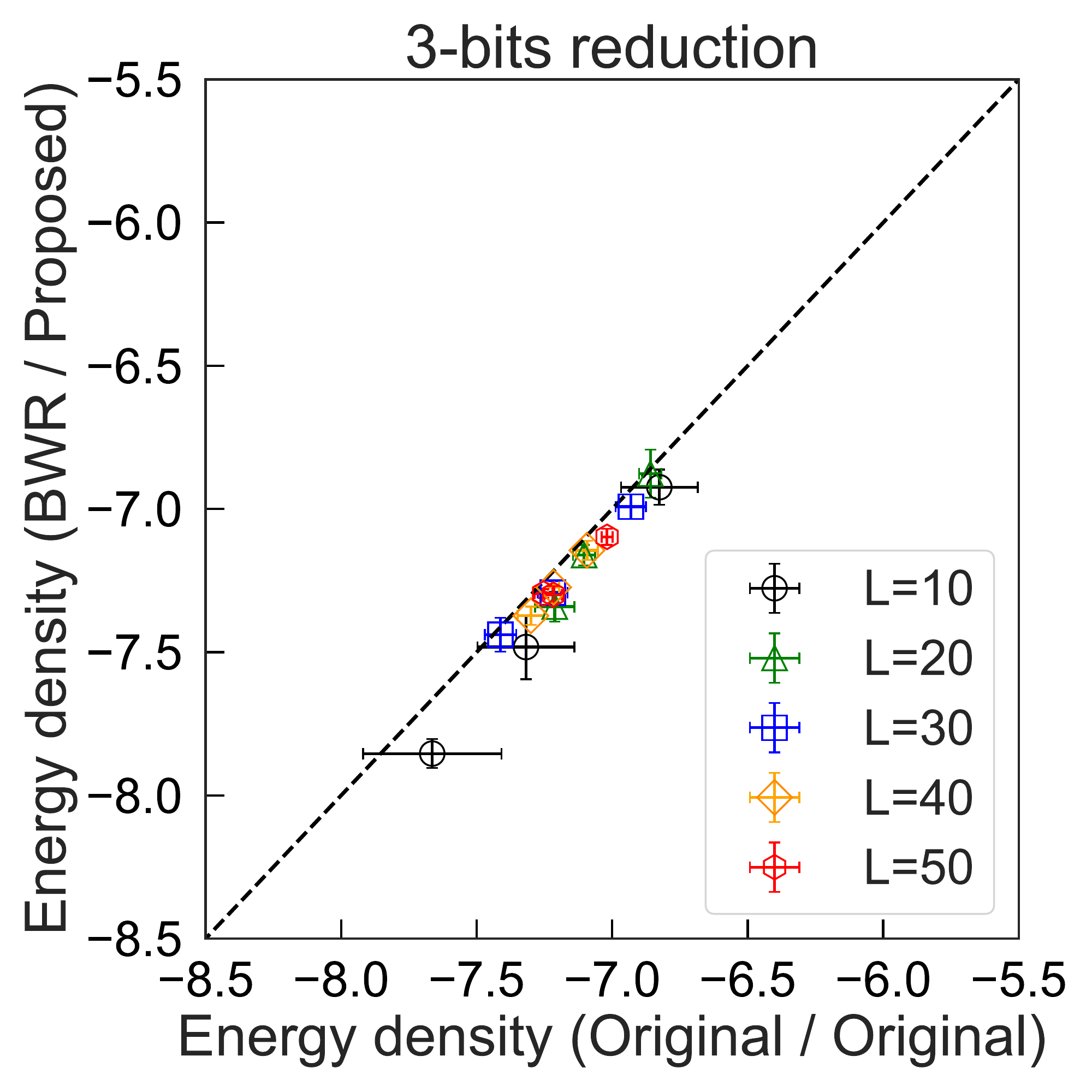}
    \label{fig:scatter_3bits_proposed}
  }
  \subfigure[]{
    \includegraphics[clip,width=0.23\linewidth]{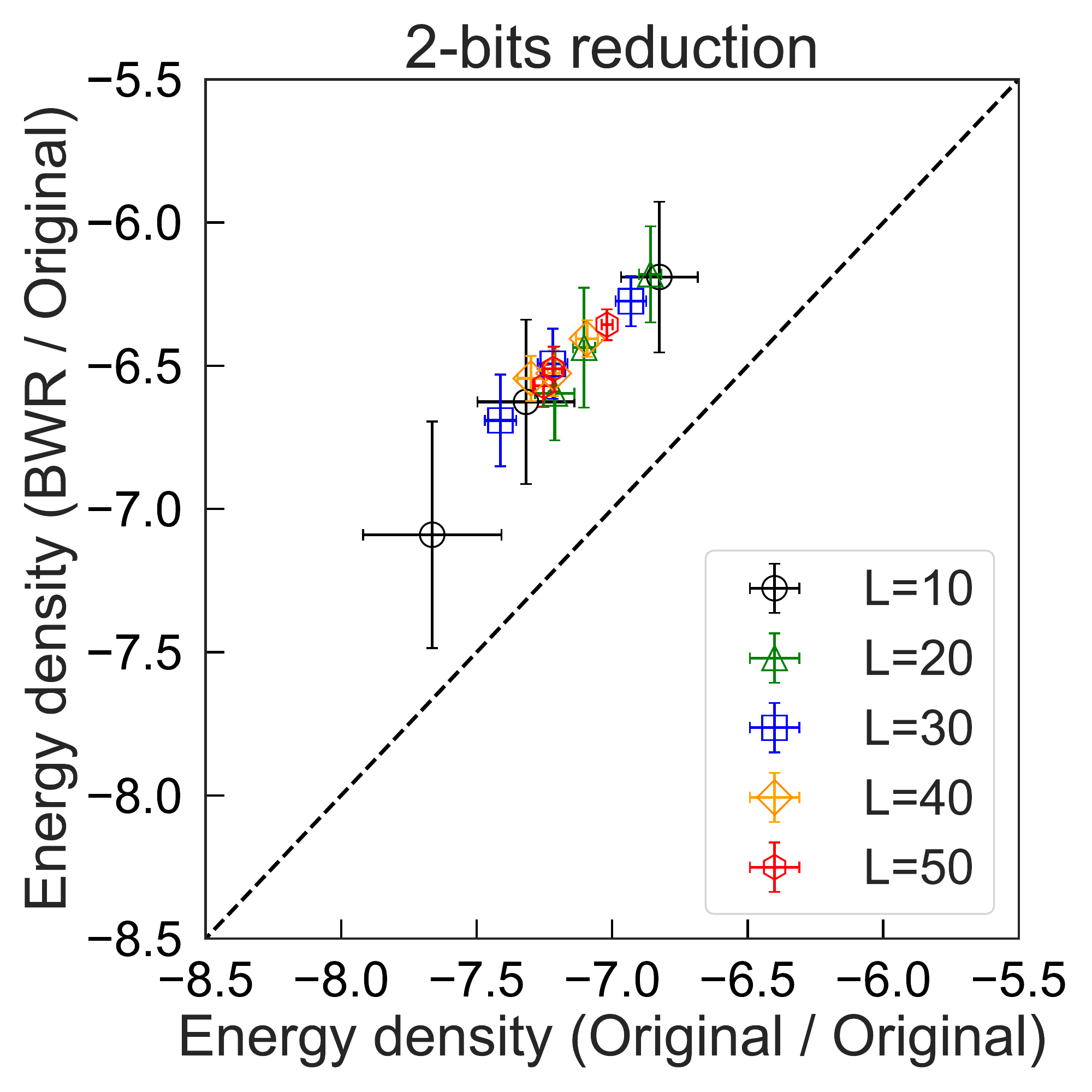}
    \label{fig:scatter_2bits_original}
  }
  \subfigure[]{
    \includegraphics[clip,width=0.23\linewidth]{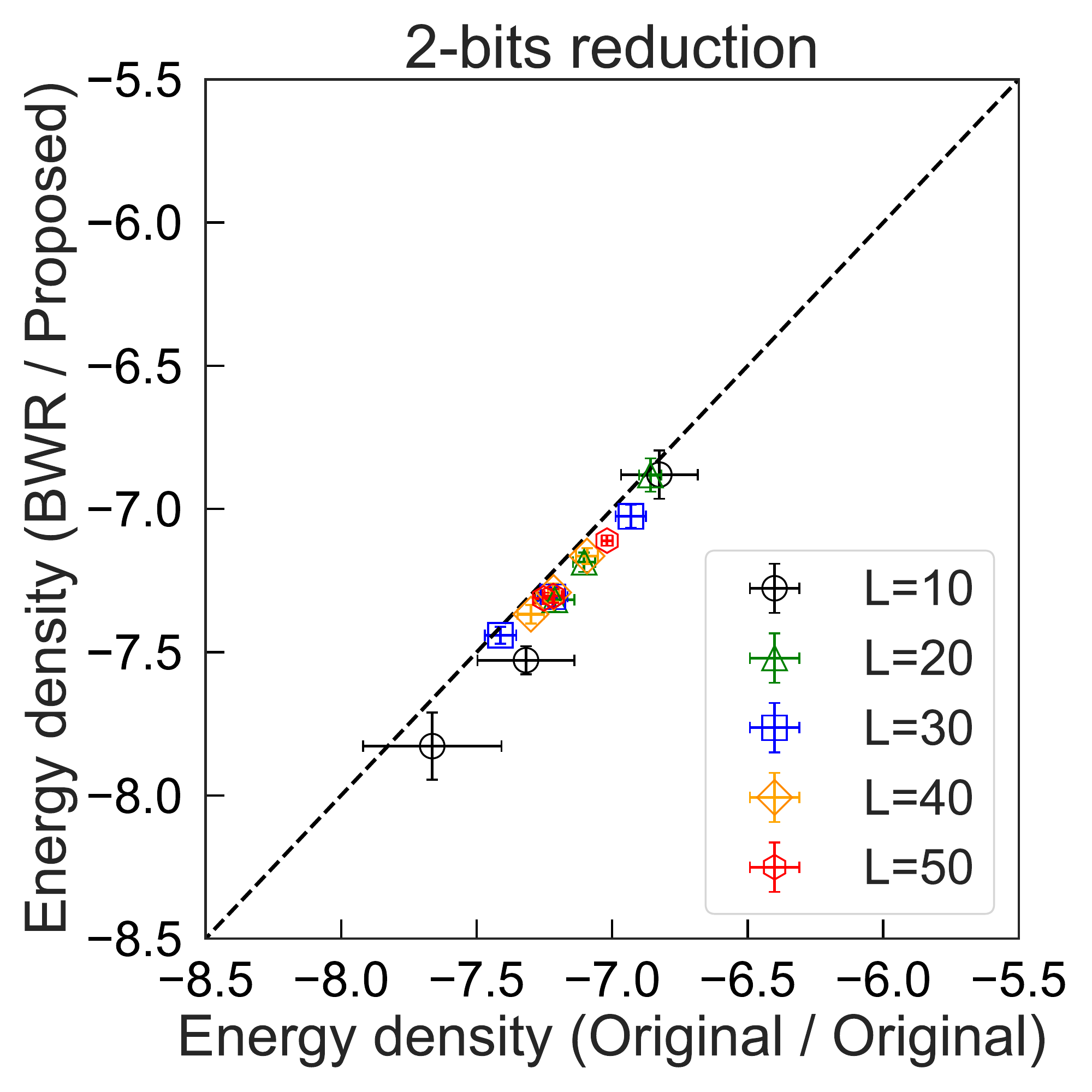}
    \label{fig:scatter_2bits_proposed}
  }
  \caption{
    Comparisons of the energy densities for several sizes of Ising models between the BWR Ising model and the original Ising model with the original SA parameters when the bit-widths are reduced to (a) $3$-bits or (c) $2$-bits, or the BWR Ising model with the proposed SA parameters and the original Ising model with the original SA parameters when the bit-widths are reduced to (b) $3$-bits or (d) $2$-bits. Every plot is the average of ten runs. The error bars are standard deviations.
  }
  \label{fig:scatter}
\end{figure*}

Next, we performed SA with several sizes of Ising models~\cite{instances} to evaluate the problem size dependency of the proposed SA parameters.
We set $L=10, 20, 30, 40$, or $50$ as the number of system spins.
The coefficients and SA parameters are as described above.
Fig.~\ref{fig:scatter} compares the energy densities for each Ising model with a different size.
Figs.~\ref{fig:scatter_3bits_original} and~\ref{fig:scatter_2bits_original} compare the energy densities at the end of SA between the original and BWR Ising models using the original SA parameters.
All points are plotted above the diagonal, indicating that the BWR Ising models have an inferior solution accuracy compared to the original Ising model when using the original SA parameters. 
Figs.~\ref{fig:scatter_3bits_proposed} and~\ref{fig:scatter_2bits_proposed} compare the energy densities at the end of SA between the original Ising model with the original SA parameters and the BWR Ising model with the proposed SA parameters.
All points are plotted below or on the diagonal, indicating that the BWR Ising model is the same or superior to the original Ising model when using the proposed SA parameters.
These results imply that this feature is independent of the system size, at least for the range considered in this study.
Additionally, it was demonstrated that the dynamical process is also independent of the system size (Appendix~\ref{sec:appendixB}).
These results suggest that our proposed SA parameters exhibit robustness to spin size in the square lattice system.

\section{Discussion}
\label{sec:Discussion}
The dynamical process of the BWR Ising model with the proposed SA parameters is almost the same as that of the original Ising model, although the early stages of the iterations differ when reduced to $3$-bits (Fig.~\ref{fig:dynamics_random}).
The difference is attributed to the coefficient used as the basis for modifying the SA parameters.
In the previous section, we used the maximum absolute values of the original Ising model as the basis.
However, there is a gap between the temperature schedule based on the maximum absolute value of the coefficient and the other coefficients (Fig.~\ref{fig:proposed_temp}).
The temperature schedule based on the maximum absolute value of the coefficient has the lowest temperature in the early stages of the outer loop iterations and the slowest temperature decrease. 

Additionally, the coefficients of the Ising model used in the previous section were generated uniformly at random. 
In many scenarios, the coefficients are not seven.
That is, the proportion of coefficients (i.e., $|J|$, $|h| = 6$ to $4$) with a large gap from the temperature schedule based on the maximum absolute value of coefficient when reduced to $3$-bits is relatively high (Fig.~\ref{fig:proposed_temp_3bits}).
This indicates that the proposed temperature schedule is set excessively low and slow for many coefficients.
Consequently, the dynamical process of the BWR Ising model in the early stages of outer loop iterations differs from that of the original Ising model. 

The solution accuracy of the BWR Ising models with the proposed SA parameters is slightly superior to that of the original Ising model (Figs.~\ref{fig:scatter_3bits_proposed} and~\ref{fig:scatter_2bits_proposed}).
This is attributed to the coefficient used as the basis for modifying the SA parameters.
The effective relaxation time $\tau_\textrm{eff}$ based on the maximum absolute value of the coefficient is larger than that based on the other coefficients at a low temperature (Fig.~\ref{fig:proposed_inner}).
Because the proposed inner loop is set by an excess $\tau_\textrm{eff}$ based on an excessively low and slow temperature schedule, for many coefficients, the probability of reaching thermal equilibrium at each temperature is higher and the solution accuracy is improved. 

\section{Conclusion and future work}
\label{sec:conclusion}

The dynamical process with SA is compared between the original Ising model and the BWR Ising model by applying the proposed method using square lattice systems.
Because the dynamical process of the BWR Ising model significantly differs from that of the original Ising model, we analyzed the BWR Ising model from the viewpoint of statistical mechanics.
The BWR Ising model with the addition of auxiliary spins has two-characteristic properties not present in the original Ising model: an effective temperature and a slow relaxation.
Considering the analytical results, we proposed SA parameters for the BWR Ising model.
Our results demonstrate that the dynamical processes of the BWR Ising model with the proposed SA parameters are close to that of the original Ising model. 

We expect that the bit-width reduction method and our parameter modification method will effectively solve the Ising model on an implemented Ising machine with a bit-width limitation of coefficients.
However, these methods are not efficient in terms of computation time.
The computation time increases as the number of auxiliary spins increases.
The number of auxiliary spins $N_\mathrm{a}$ to be added by the proposed method per one coefficient is represented by:
\begin{align}
  N_\mathrm{a} \simeq 
  \begin{cases}
    2^{\lparen n_\mathrm{original}-n_\mathrm{BWR}\rparen} & (n_\mathrm{BWR}>2)\\
    2^{\lparen n_\mathrm{original} - 1\rparen}-2 & (n_\mathrm{BWR}=2),
  \end{cases}
  \label{eq:num_auxiliary_spins}
\end{align}
where $n_\textrm{original}$ and $n_\textrm{BWR}$ are the bit-widths of the original Ising model and the BWR Ising model, respectively.
The increased computation time is directly related to the inner loop.
The modified inner loop is set to $1$ MCS$\times \tau_\textrm{eff}$.
The MCS increases with the number of auxiliary spins because MCS is the total number of spins of the Ising model.
$\tau_\textrm{eff}$ increases with the number of auxiliary spins and temperature.
The modified temperature schedule is lower than that of the original temperature schedule due to the increased number of auxiliary spins.
Therefore, the computation time increases. 

A method has been proposed that combines the shift method with the bit-width reduction method using auxiliary spins, to reduce the number of auxiliary spins required~\cite{yachi2023efficient}.
However, this approach leads to a different ground state than the original Ising model.
One alternative method to maintain the ground state while mitigating the increase in computation time is to expand the bit-width that can be input to the Ising machine.
However, even if the bit-width of the Ising machine cannot be increased due to hardware limitations, the number of the MCS can be reduced if the auxiliary spins can be flipped simultaneously similar to CMOS annealing~\cite{Okuyama2017}.
This approach should be considered in the future.
Additionally, we plan to investigate other implemented algorithms of Ising machines such as quantum annealing~\cite{tanaka2010nonmonotonic}. 

\appendices
\section{Analysis of the entropic effects caused by auxiliary spin}
\label{sec:appendixA}

This appendix provides detailed derivations of the definitions for the effective magnetic field $L_\textrm{eff}$ and the effective interaction $K_\textrm{eff}$ (see~(\ref{eq:Leff_def}) in the main text).

\subsubsection{Magnetic fields}
\label{subsec:Leff}
First, consider the case of a bit-width reduction for a magnetic field. 
The Ising model is assumed to be the same as that in Section~\ref{subsec:magnetic_fields} (Fig.~\ref{fig:red_method_mag}).
To derive the definition of the effective magnetic field, the expectation value of the system spin $\langle\sigma_1\rangle$ of the original Ising model and that of the BWR Ising model are matched.
The expectation value of the original Ising model $\langle\sigma_1\rangle_\textrm{original}$ is given by
\begin{align}
  \langle\sigma_1\rangle_\mathrm{original}=\sum_{\sigma_1=\pm1}\sigma_1 \times P=\frac{\sum_{\sigma_1=\pm1}\sigma_1e^{L{\sigma_1}}}{\sum_{\sigma_1=\pm1}e^{L{\sigma_1}}},
  \label{eq:mg_exp_original}
\end{align}
where $L=h/T$ and $P$ is the probability distribution at temperature $T$.
$P$ is given by
\begin{align}
  P=\frac{e^{-\beta H}}{\sum_{\sigma_i=\pm1}e^{-\beta H}}.
  \label{eq:probability_distribution}
\end{align}

The expectation value of the BWR Ising model$\langle\sigma_1\rangle_\textrm{BWR}$ is similarly given by
\begin{align}
    \langle\sigma_1\rangle_\mathrm{BWR}=\frac{\sum_{\sigma_1=\pm1}\sigma_1e^{L_\mathrm{eff}{\sigma_1}}}{\sum_{\sigma_1=\pm1}e^{L_\mathrm{eff}{\sigma_1}}},
  \label{eq:mg_exp_BWR}
\end{align}
where $L_\textrm{eff}=h/T_\textrm{eff}$ and $T_\textrm{eff}$ is the effective temperature. 
To match~(\ref{eq:mg_exp_original}) and~(\ref{eq:mg_exp_BWR}), the auxiliary spins in~(\ref{eq:mg_exp_BWR}) are partially summed to obtain the marginal probability for $\sigma_1$.
Therefore, the effective magnetic field $L_\textrm{eff}$ is defined by tracing the auxiliary spins $s_i$ to give~(\ref{eq:Leff_def}) in the main text.

\subsubsection{Interactions}
\label{subsec:Keff}
In the case of a bit-width reduction of the interactions, the Ising model shown in Fig.~\ref{fig:red_method_int} is assumed.
The expectation values of the original Ising model $\langle\sigma_1\sigma_2\rangle_\textrm{original}$ and the BWR Ising model $\langle\sigma_1\sigma_2\rangle_\textrm{BWR}$ are given by
\begin{align}
  \langle\sigma_1\sigma_2\rangle_\mathrm{original}=\frac{\sum_{\sigma_i=\pm1}\sigma_1\sigma_2e^{K{\sigma_1\sigma_2}}}{\sum_{\sigma_i=\pm1}e^{K{\sigma_1\sigma_2}}},
  \label{eq:int_exp_original}
\end{align}
\begin{align}
  \langle\sigma_1\sigma_2\rangle_\mathrm{BWR}=\frac{\sum_{\sigma_i=\pm1}\sigma_1\sigma_2e^{K_\mathrm{eff}{\sigma_1\sigma_2}}}{\sum_{\sigma_i=\pm1}e^{K_\mathrm{eff}{\sigma_1\sigma_2}}},
  \label{eq:int_exp_BWR}
\end{align}
where $K=J/T$ and $K_\textrm{eff}=J/T_\textrm{eff}$, respectively.
To match~(\ref{eq:int_exp_original}) and~(\ref{eq:int_exp_BWR}), the effective interaction $K_\textrm{eff}$ is defined by tracing the auxiliary spins $s_i$, which yields~(\ref{eq:Keff_def}) in the main text.

\section{Dynamical process for large-size square lattice system}
\label{sec:appendixB}

In this appendix, to evaluate the problem size dependency of the dynamical process, we performed SA on a large-size square lattice system ($L=40, 50$)~\cite{instances} with several types of SA parameters.

\begin{figure}[t]
  \centering
  \subfigure[]{
    \includegraphics[clip,width=0.465\linewidth]{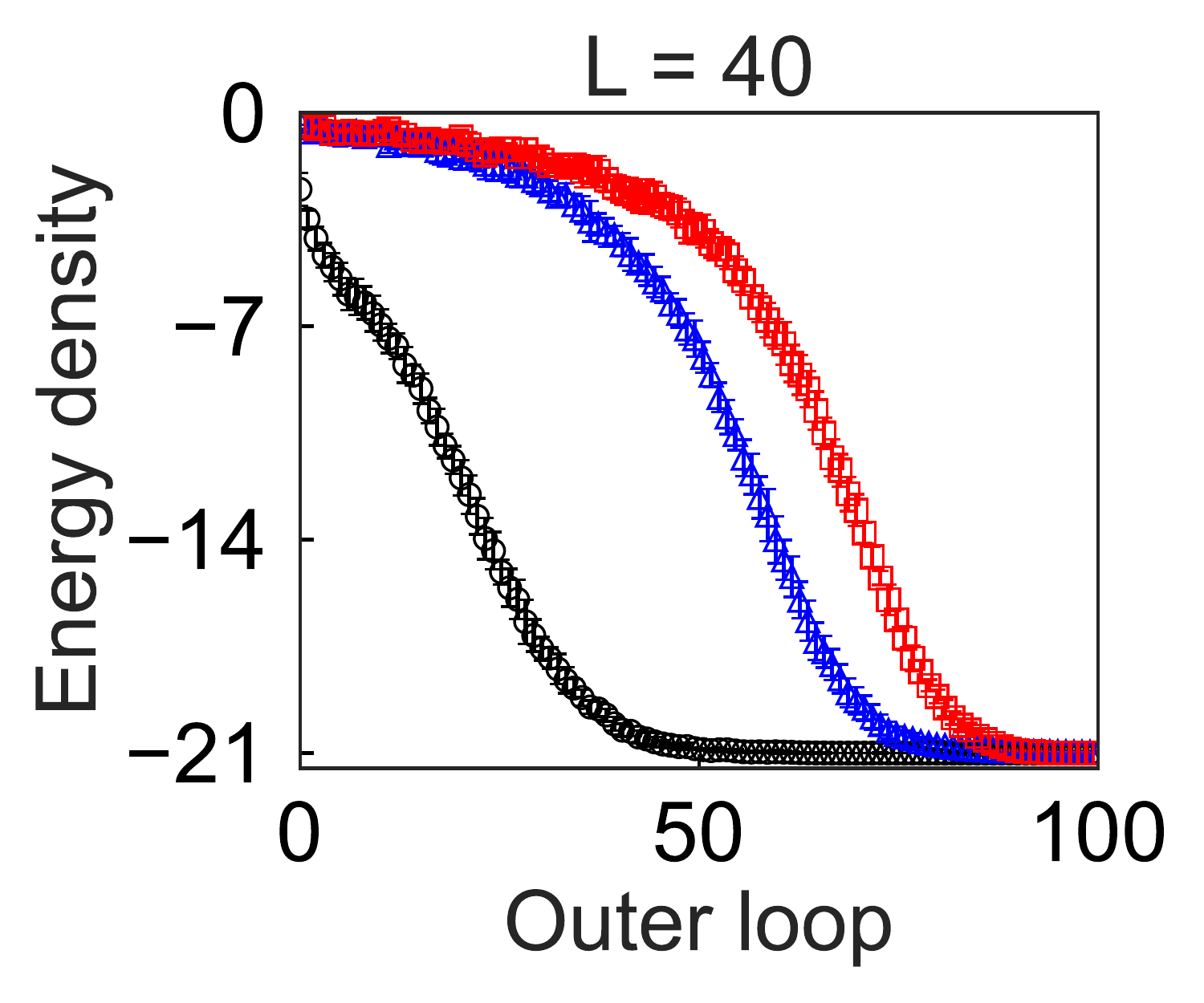}
    \label{fig:L40o}
  }
  \subfigure[]{
    \includegraphics[clip,width=0.465\linewidth]{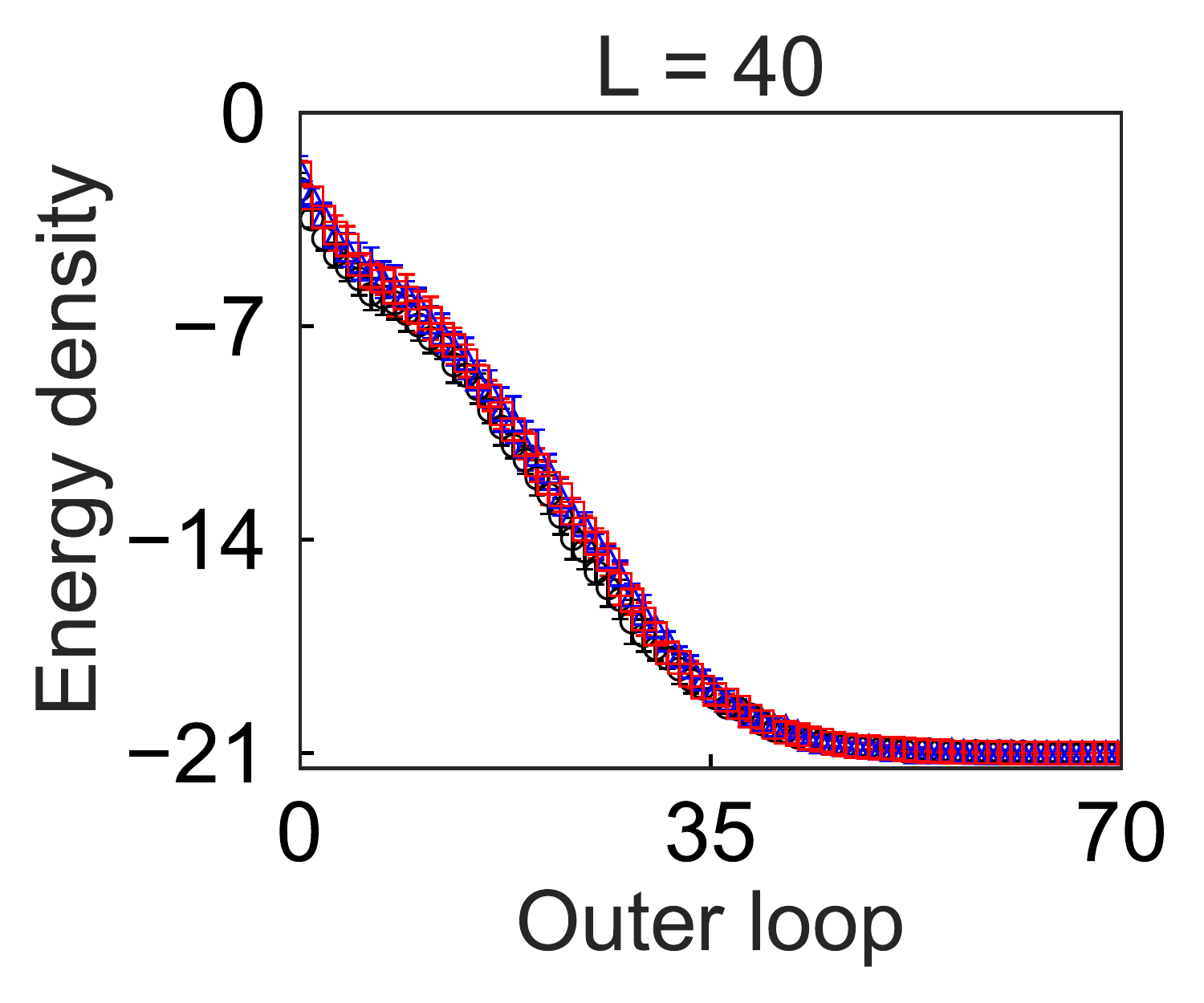}
    \label{fig:L40p}
  }
  \\
  \subfigure[]{
    \includegraphics[clip,width=0.465\linewidth]{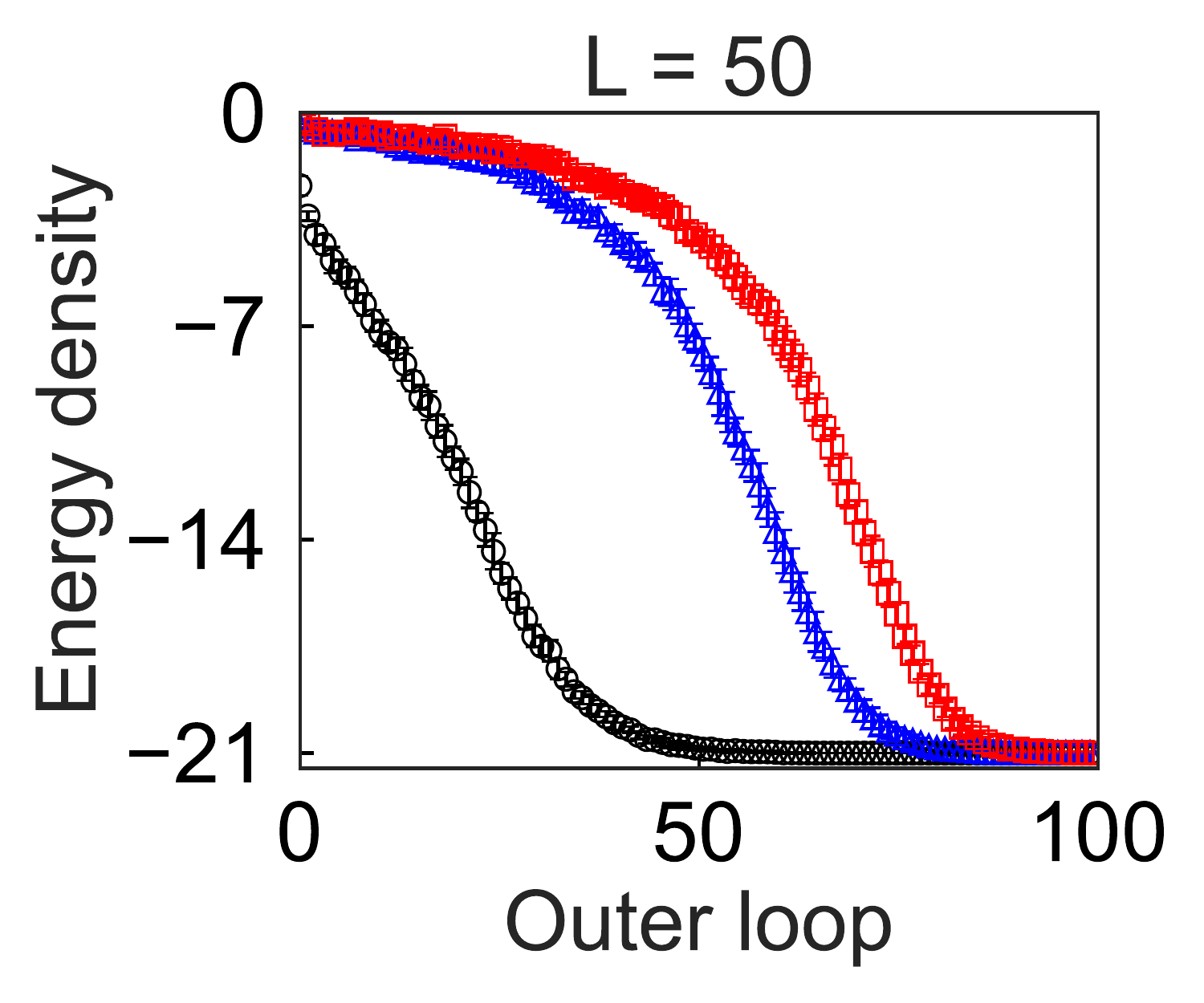}
    \label{fig:L50o}
  }
  \subfigure[]{
    \includegraphics[clip,width=0.465\linewidth]{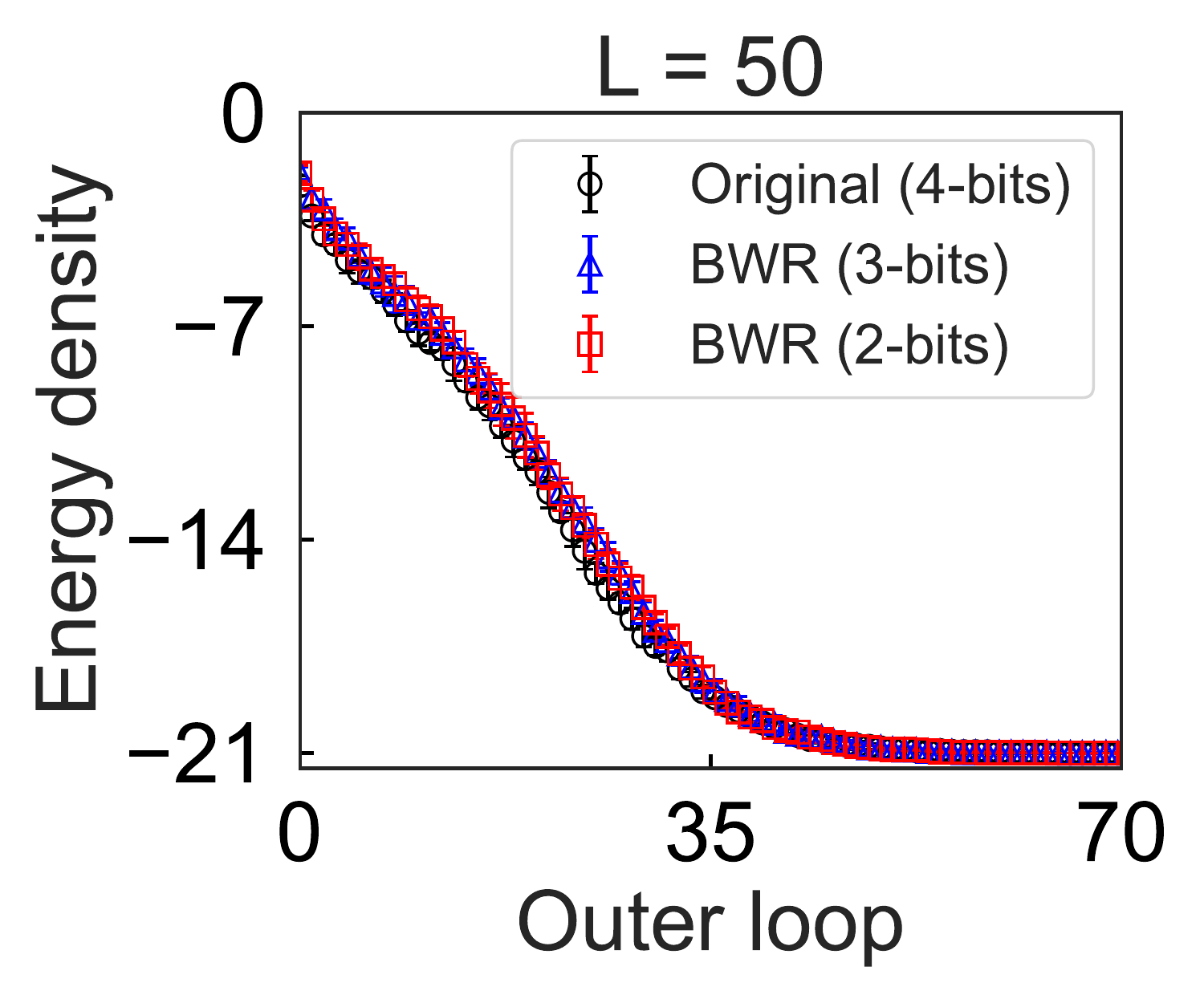}
    \label{fig:L50p}
  }
  \caption{
    Dynamical processes of the original Ising model with the original SA parameters and the BWR Ising model with (a), (c) the original SA parameters, or (b), (d) the proposed SA parameters. Square lattice system size are (a), (b) $L=40$ and (c), (d) $L=50$. The BWR Ising model ($2$-bits), the BWR Ising model ($3$-bits) and the original Ising model ($4$-bits) are denoted by red squares, blue triangles and black circles. Every plot is an average of ten runs. The error bars are standard deviations.
  }
  \label{fig:simulation_large}
\end{figure}

We first investigated the dynamical process of the square lattice system with all coefficients set to 7, as described in Section~\ref{sec:simulation}.
The dynamical processes are shown in Fig.~\ref{fig:simulation_large}.
The results revealed that even in large-size square lattice systems, the dynamical processes of the original Ising model and the BWR Ising model differ when using the original SA parameters, shown in Table~\ref{table:parameters_simulation}.
However, by using the proposed SA parameters described in Section~\ref{sec:proposed_parameter}, the dynamical processes of the original Ising model and the BWR Ising model became almost the same.

\begin{figure*}[t]
  \centering
  \subfigure[]{
    \includegraphics[clip,width=0.4\linewidth]{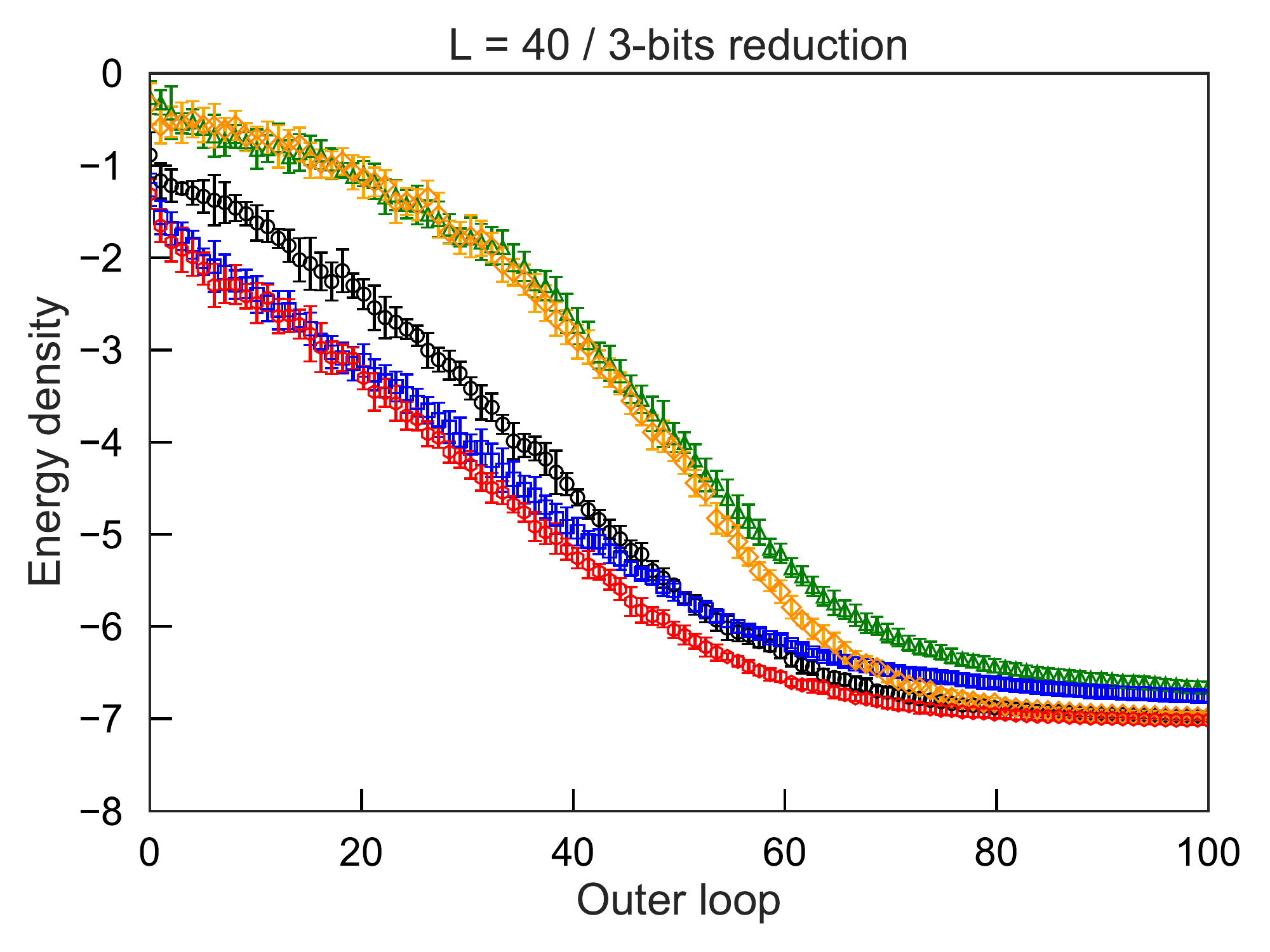}
    \label{fig:d3_40}
  }
  \subfigure[]{
    \includegraphics[clip,width=0.4\linewidth]{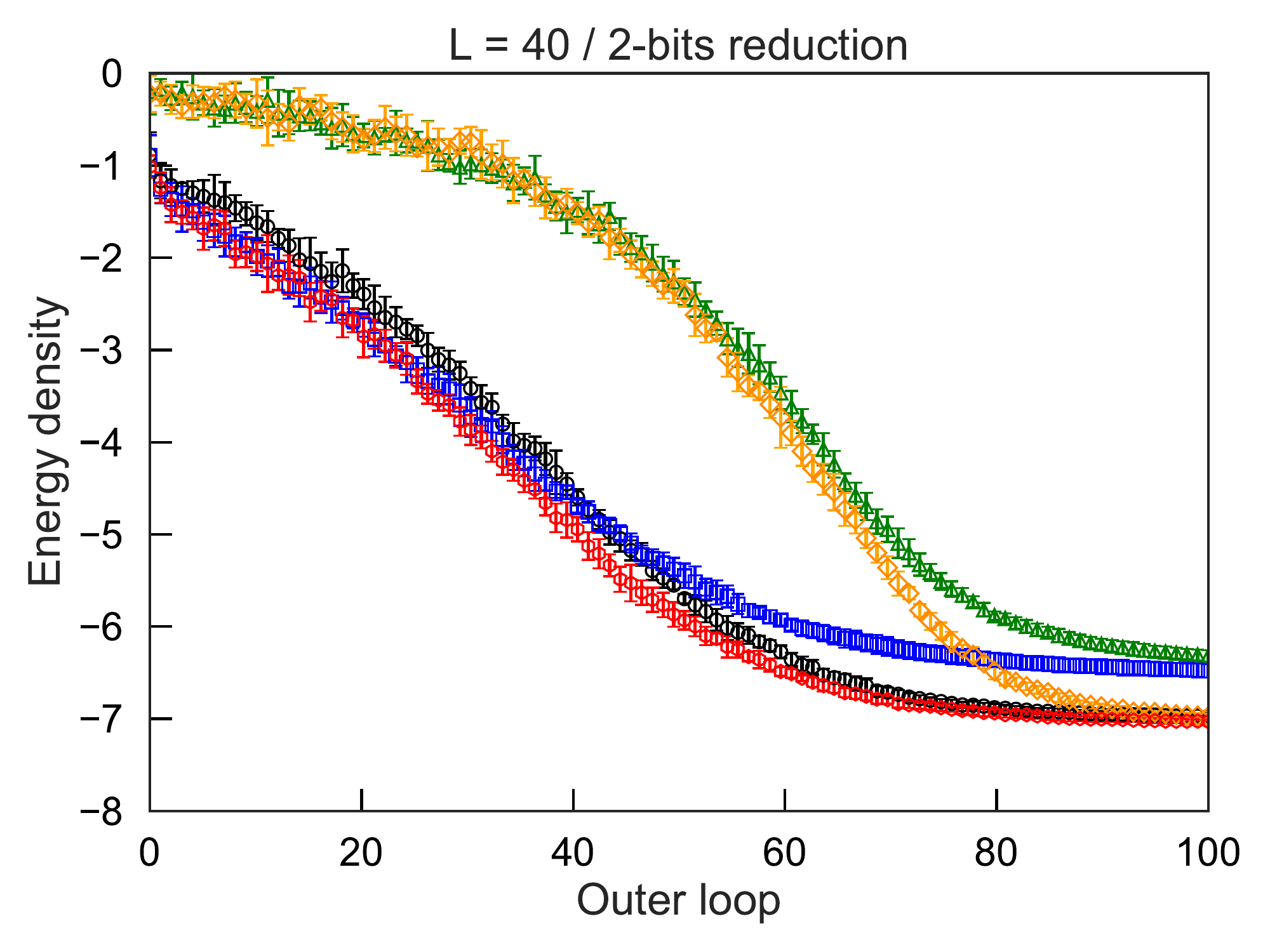}
    \label{fig:d2_40}
  }
  \\
  \subfigure[]{
    \includegraphics[clip,width=0.4\linewidth]{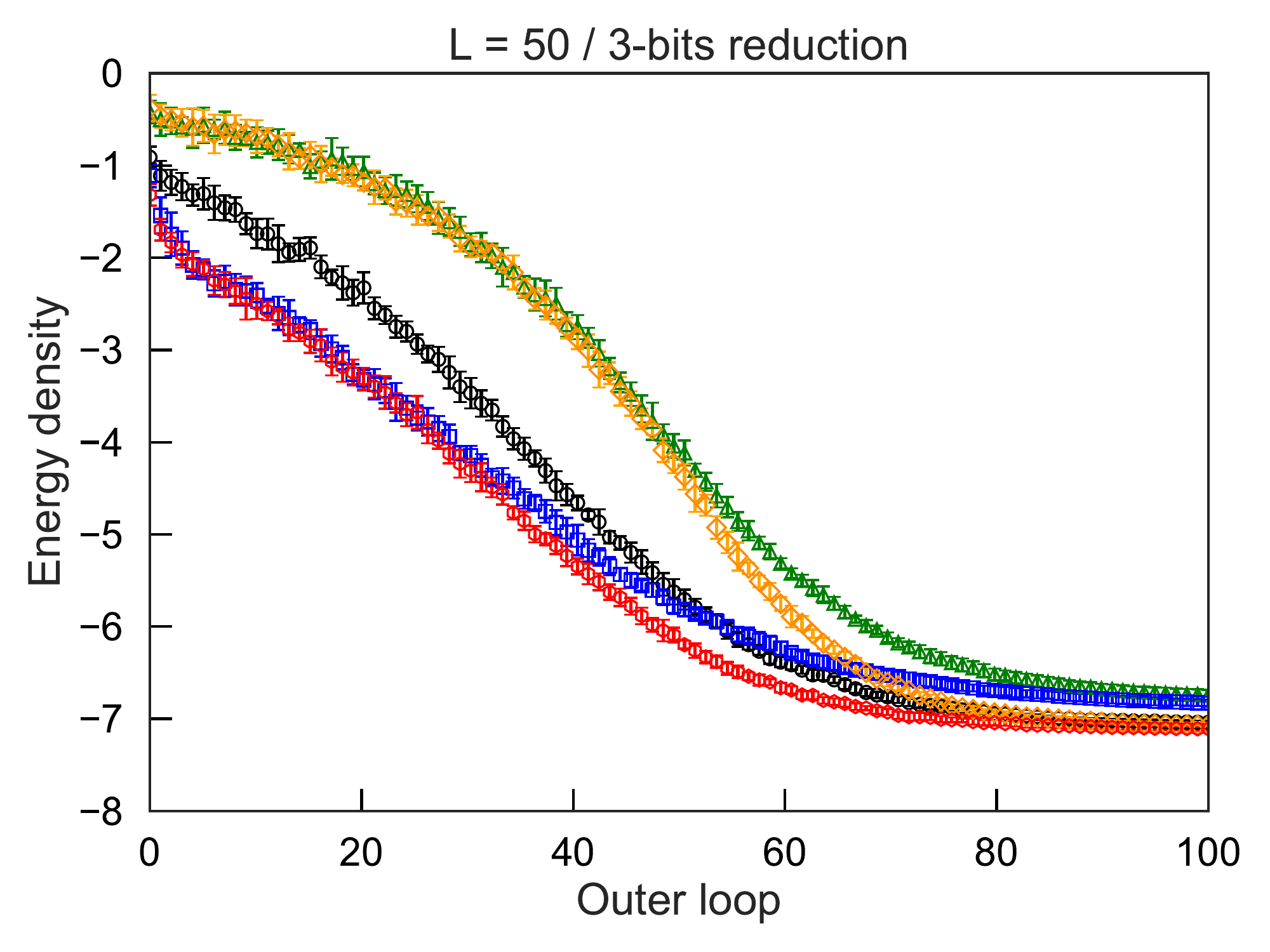}
    \label{fig:d3_50}
  }
  \subfigure[]{
    \includegraphics[clip,width=0.4\linewidth]{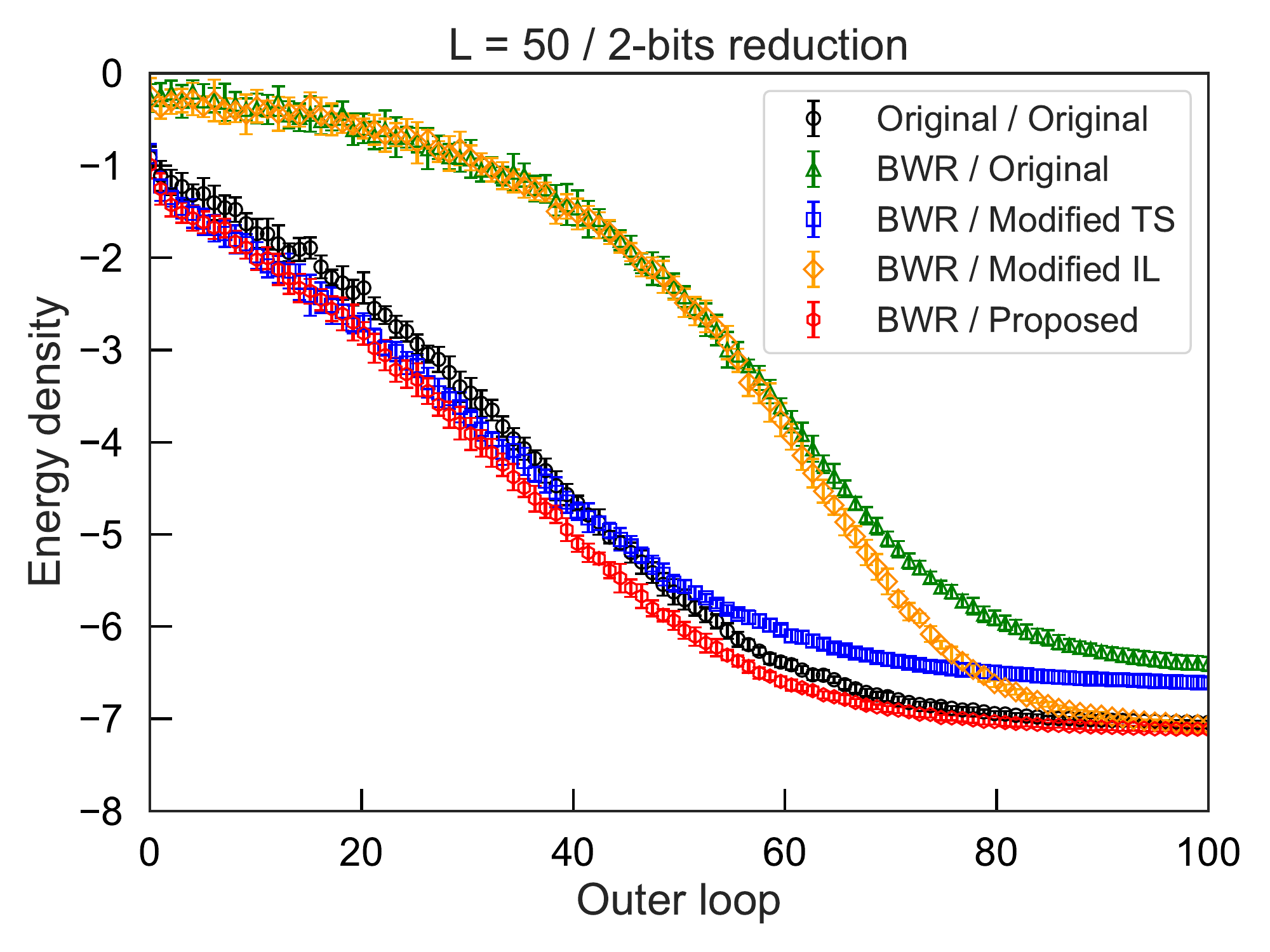}
    \label{fig:d2_50}
  }
  \caption{
    Dynamical processes of the original Ising model and the BWR Ising model reduced to (a), (c) $3$-bits or (b), (d) $2$-bits. Square lattice system sizes are (a), (b) $L=40$ and (c), (d) $L=50$. Black circles denote the original Ising model. Green triangles, blue squares, orange diamonds, and red hexagons denote the BWR Ising model with the original SA parameters, modified temperature schedule only, modified inner loop only, and proposed SA parameters, respectively. Every plot is the average of ten runs. The error bars are standard deviations.
  }
  \label{fig:dynamics_large}
\end{figure*}

Next, we investigated the dynamical process of a large-size square lattice system with randomly assigned coefficients, as described in Section~\ref{sec:random_ising}.
We also used the SA parameters as described in the same Section~\ref{sec:random_ising}.
Figure~\ref{fig:dynamics_large} shows the dynamical processes.
As mentioned in Section~\ref{sec:random_ising}, the dynamical processes of the BWR Ising model with the proposed SA parameters became closer to that of the original Ising model.

\section{Performance of the proposed method for different temperature schedules}
\label{sec:appendixC}

\begin{figure}[t]
  \centering
  \includegraphics[clip,width=0.9\linewidth]{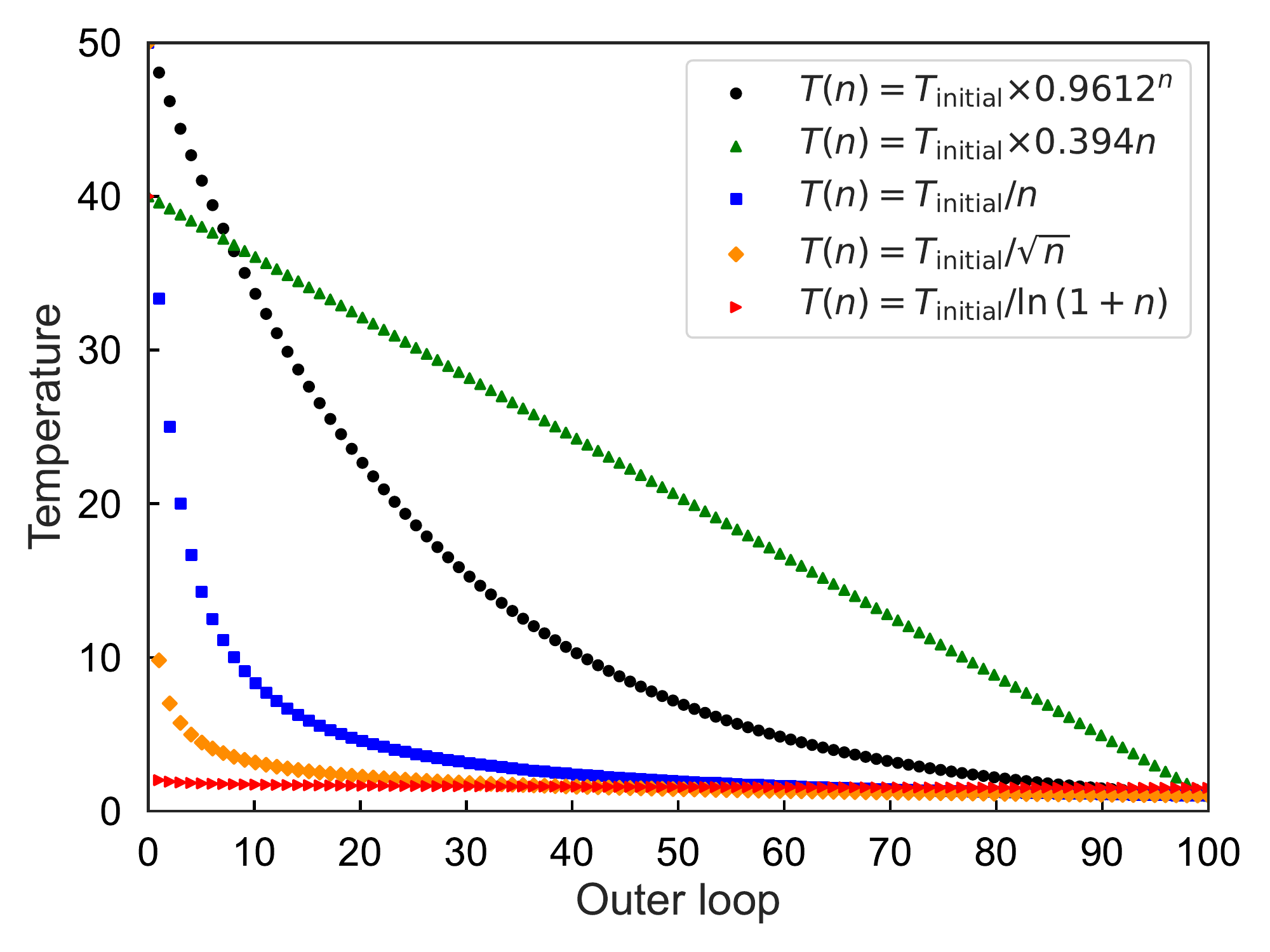}
  \caption{Different types of temperature schedules. The $n$ in the temperature schedule represents the $n$-th outer loop.}
  \label{fig:temperature_schedule}
\end{figure}

This appendix evaluates the performance of the proposed SA parameters when SA is performed at various temperature schedules.
Table~\ref{table:various_parameters} and Fig.~\ref{fig:temperature_schedule} show the temperature schedules.
The initial state, outer loop, and inner loop were set to random, $100$ ($n=0-99$), and 1 MCS, respectively, as described in Sections~\ref{sec:simulation} and~\ref{sec:random_ising}. 
These parameters were adjusted so that the final temperature was around one.

\begin{table}[b]
  \centering
  \caption{Different temperature schedules in the SA. The $n$ in the temperature schedule represents the $n$-th outer loop.}
  \label{table:various_parameters}
  \begin{tabular}{lcc} \toprule
    \multicolumn{1}{c}{\raisebox{1em}{Temperature schedule}} & \shortstack[c]{Initial \\Temperature \\($T_\textrm{initial}$)} & \raisebox{0.5em}{\shortstack[c]{Final \\Temperature}}\\ \midrule
    $T(n) = T_\textrm{initial} \times 0.9612^n$ & 50 & 0.994\\
    $T(n) = T_\textrm{initial} \times 0.394n$ & 40 & 0.994\\
    $T(n) = T_\textrm{initial} / n$ & 50 & 0.990\\
    $T(n) = T_\textrm{initial} / \sqrt{n}$ & 50 & 1.005\\
    $T(n) = T_\textrm{initial} / \ln{(1+n)}$ & 40 & 1.506\\
    \bottomrule
  \end{tabular}
\end{table}

We performed SA for a square lattice system with $L=20$.
The coefficients of the magnetic fields and interactions take values under the same conditions as those described in Section~\ref{sec:random_ising}.

\begin{figure*}[t]
  \centering
  \subfigure[]{
    \includegraphics[clip,width=0.32\linewidth]{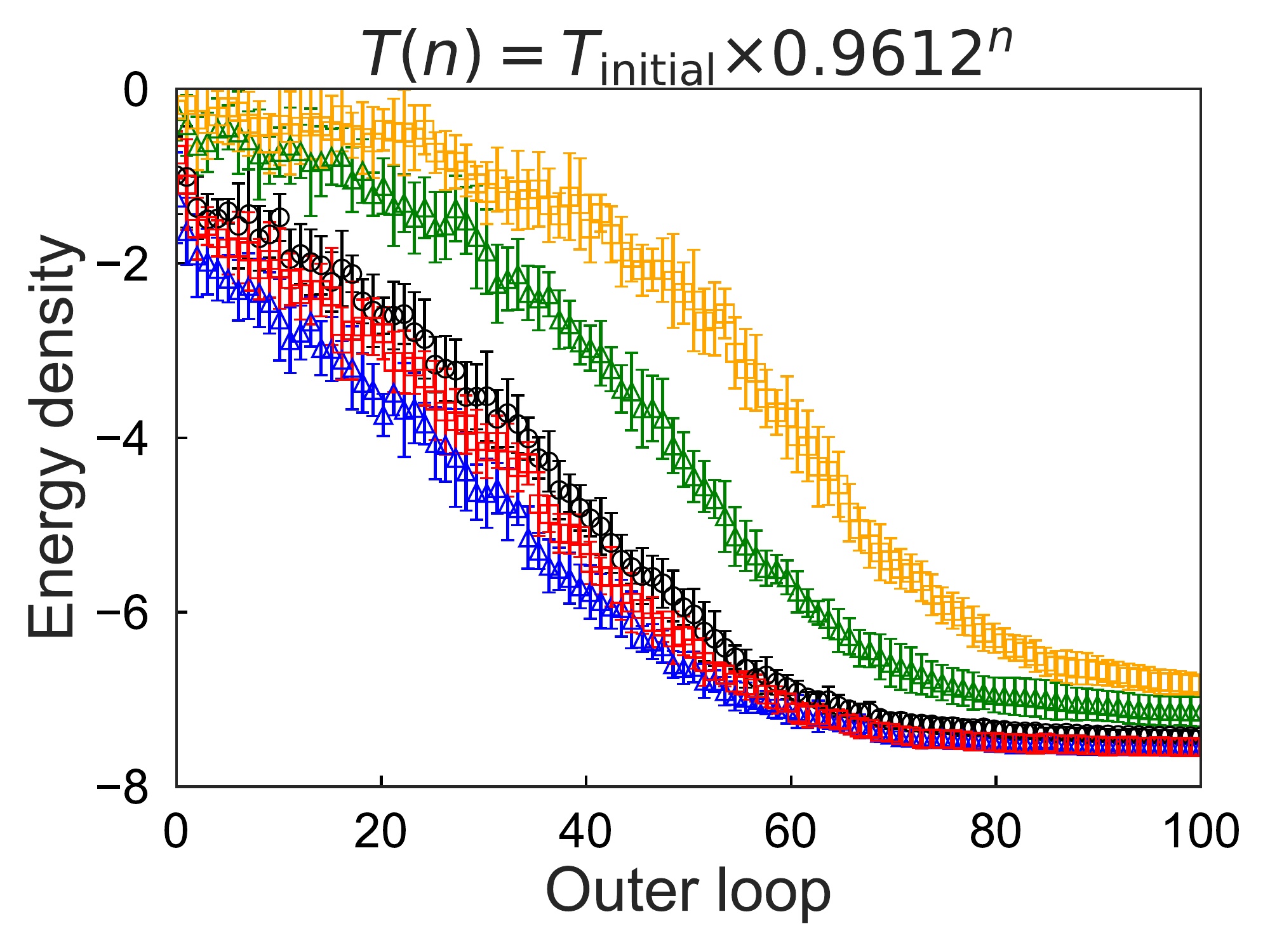}
    \label{fig:geometric}
  }
  \subfigure[]{
    \includegraphics[clip,width=0.32\linewidth]{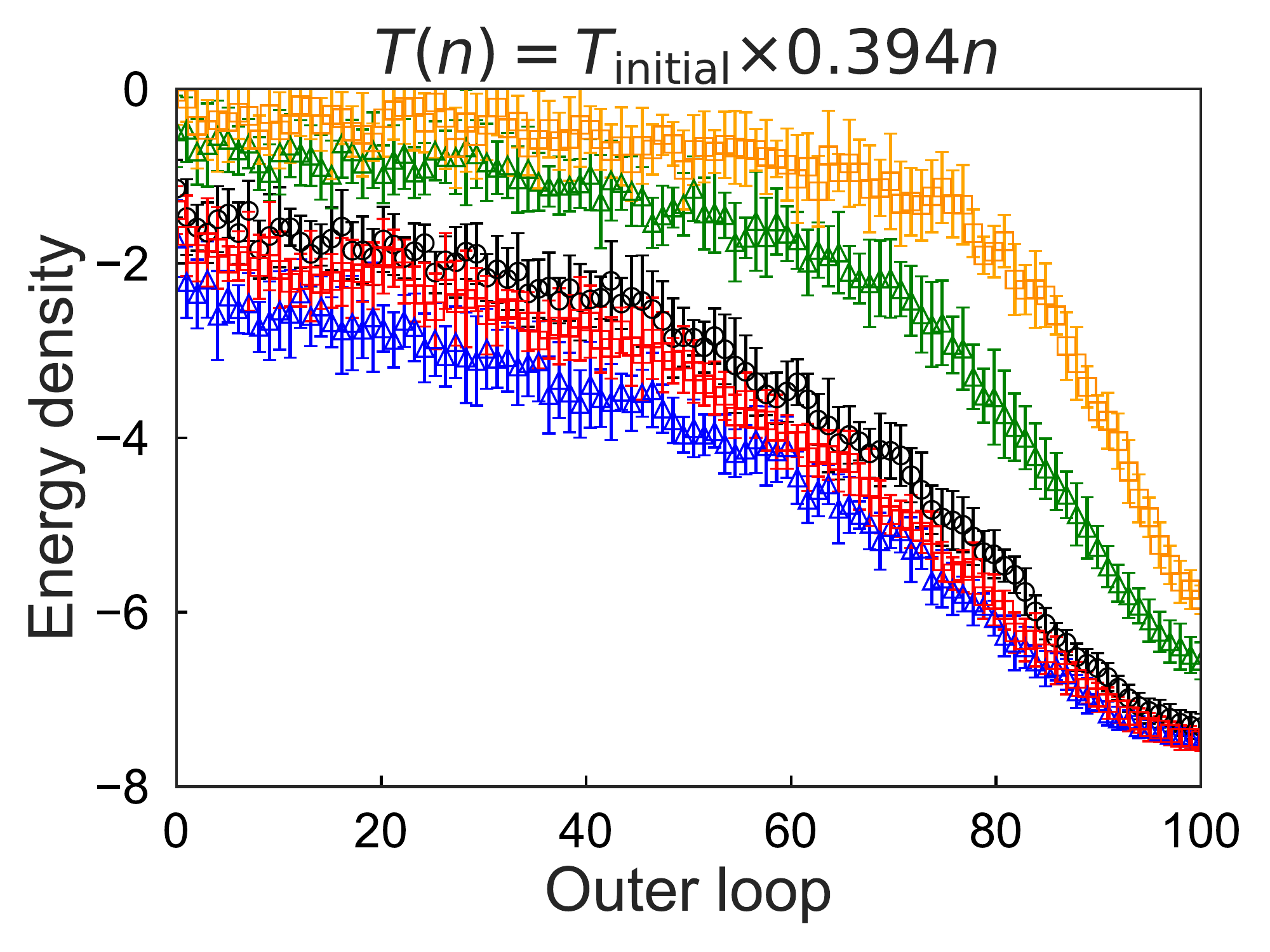}
    \label{fig:linear}
  }
  \subfigure[]{
    \includegraphics[clip,width=0.32\linewidth]{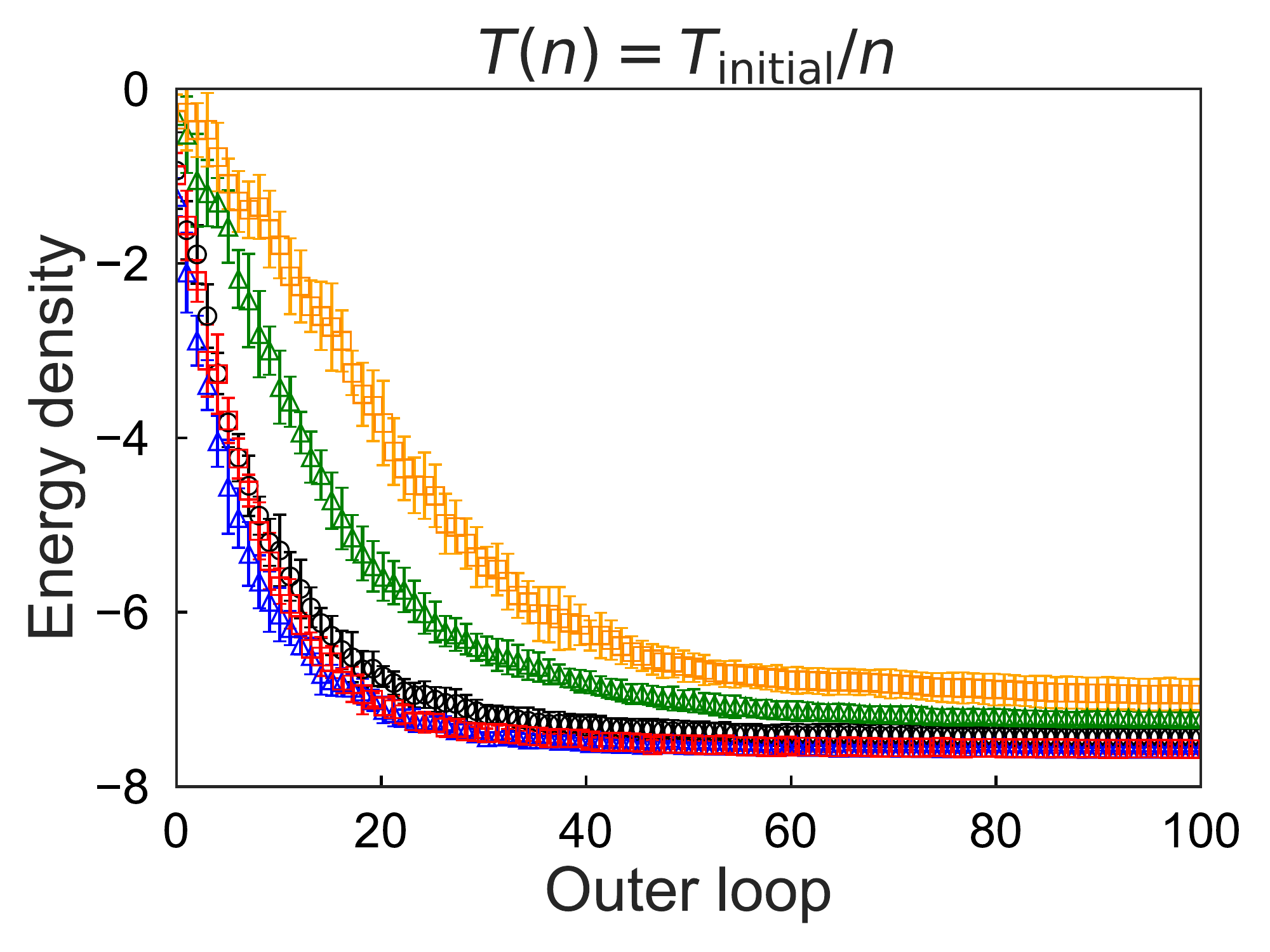}
    \label{fig:divide}
  }
  \\
  \subfigure[]{
    \includegraphics[clip,width=0.32\linewidth]{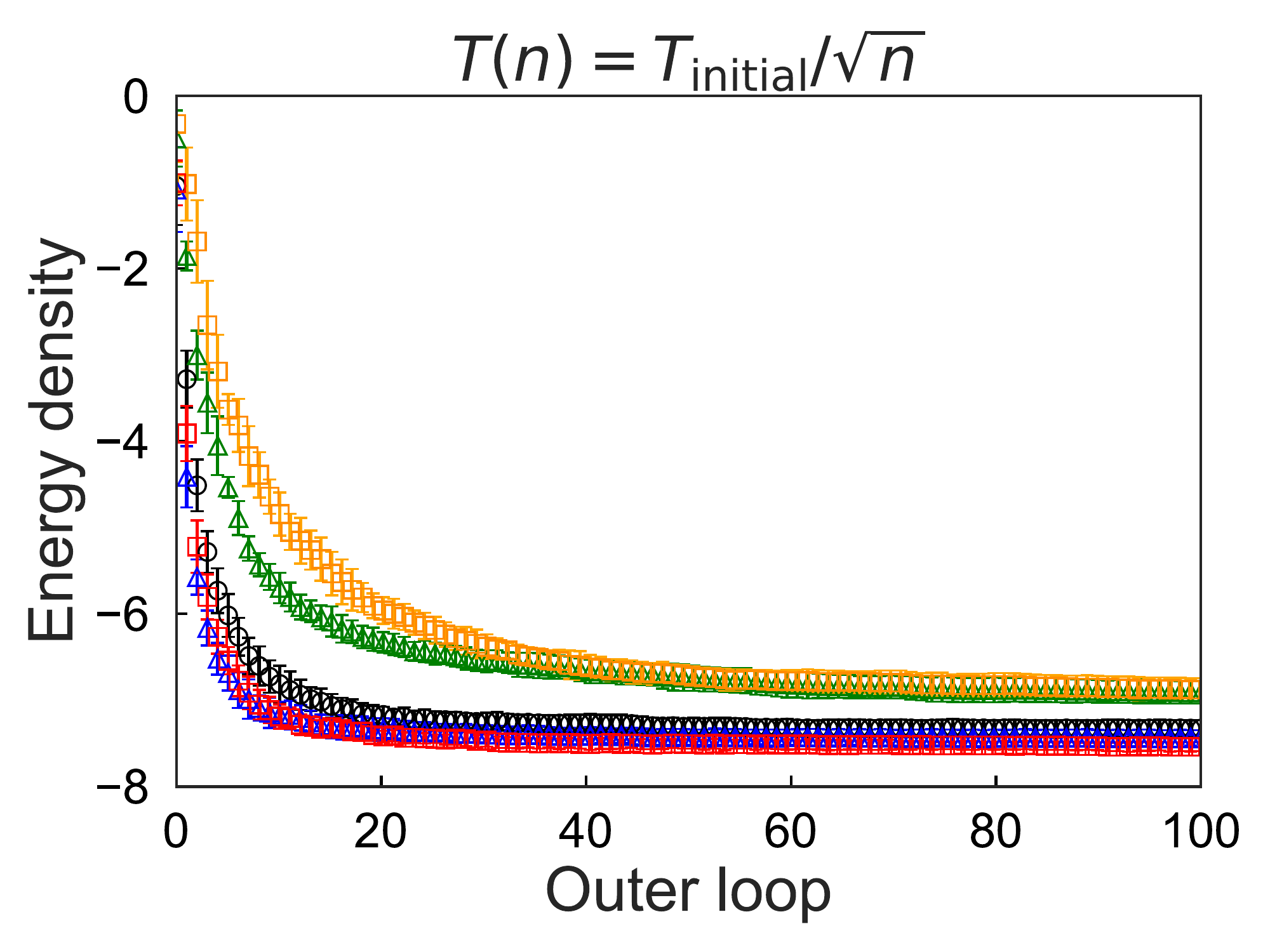}
    \label{fig:root}
  }
  \subfigure[]{
    \includegraphics[clip,width=0.32\linewidth]{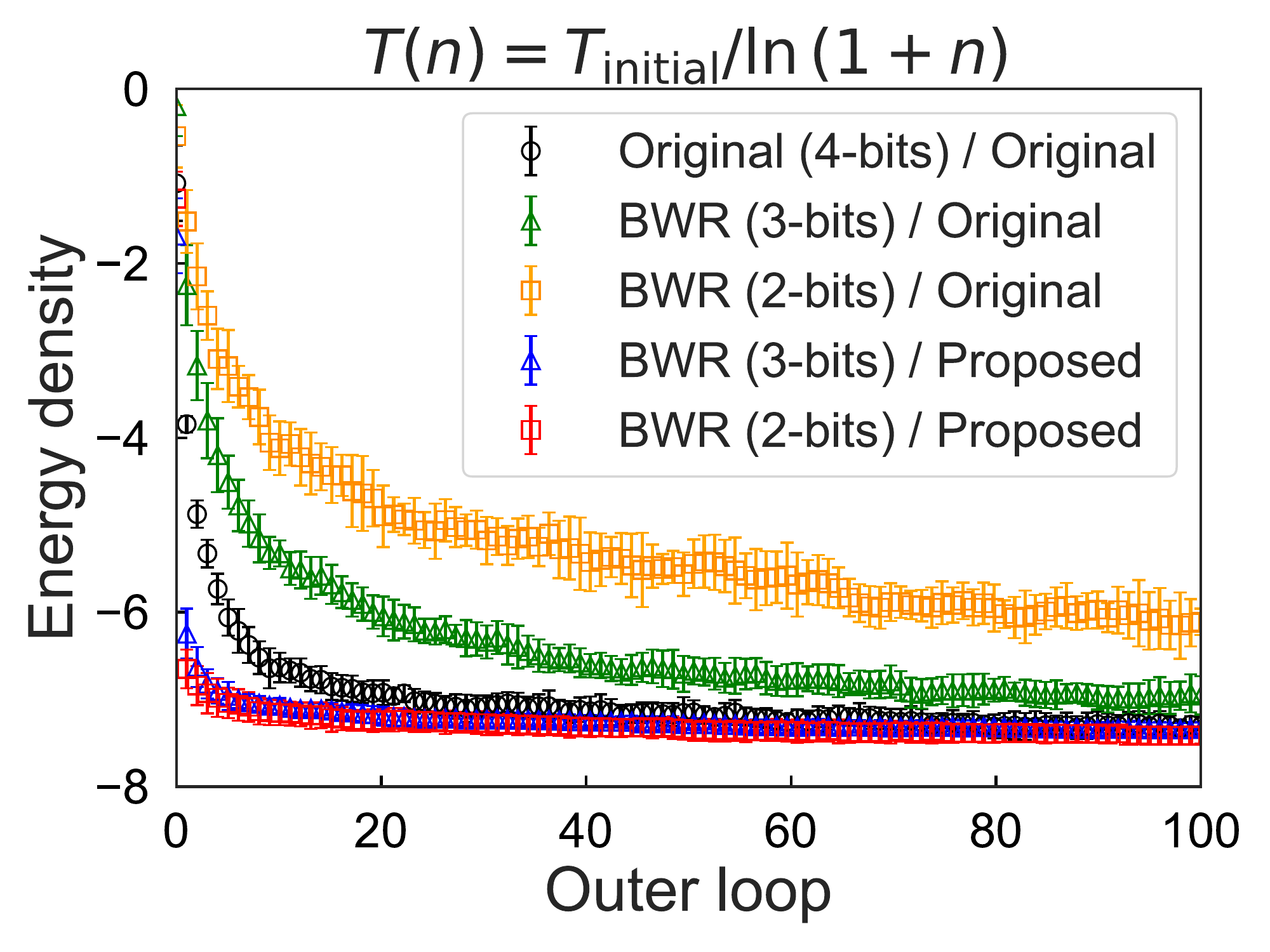}
    \label{fig:log}
  }
  \caption{
    Dynamical processes of the BWR Ising model and the original Ising model. Black circles denote the original Ising model. The $n$ in the temperature schedule represents the $n$-th outer loop. Green triangles and orange squares denote the BWR Ising model with the original SA parameters when the bit-widths are reduced to a $3$-bit width or $2$-bit width, respectively. Blue triangles and red squares denote the BWR Ising model with the proposed SA parameters when the bit-widths are reduced to $3$-bits or $2$-bits, respectively. Every plot is the average of ten runs. The error bars are standard deviations.
  }
  \label{fig:different_temperature_schedule}
\end{figure*}

Fig.~\ref{fig:different_temperature_schedule} shows the dynamical processes of the BWR Ising model with the proposed or original parameters and the original Ising model with the original parameters using a different temperature schedule.
For all temperature schedules, the dynamical processes and the energy densities at the end of SA of the BWR Ising model with the original SA parameters differed from that of the original Ising model.
In contrast, the BWR Ising model with the proposed SA parameters showed a similar performance as the original Ising model.

\section*{Acknowledgment}
This article is based on the results obtained from a project, JPNP16007, commissioned by the New Energy and Industrial Technology Development Organization (NEDO). 
The computation in this work has been partially done using the facilities of the Supercomputer Center, the Institute for Solid State Physics, the University of Tokyo.
S.~T. was supported in part JSPS KAKENHI (Grant Numbers JP21K03391, JP23H05447) and JST Grant Number JPMJPF2221.
Human Biology-Microbiome-Quantum Research Center (Bio2Q) is supported by World Premier International Research Center Initiative (WPI), MEXT, Japan.

\bibliography{reference.bib}

\begin{IEEEbiography}[{\includegraphics[width=1in,height=1.25in,clip,keepaspectratio]{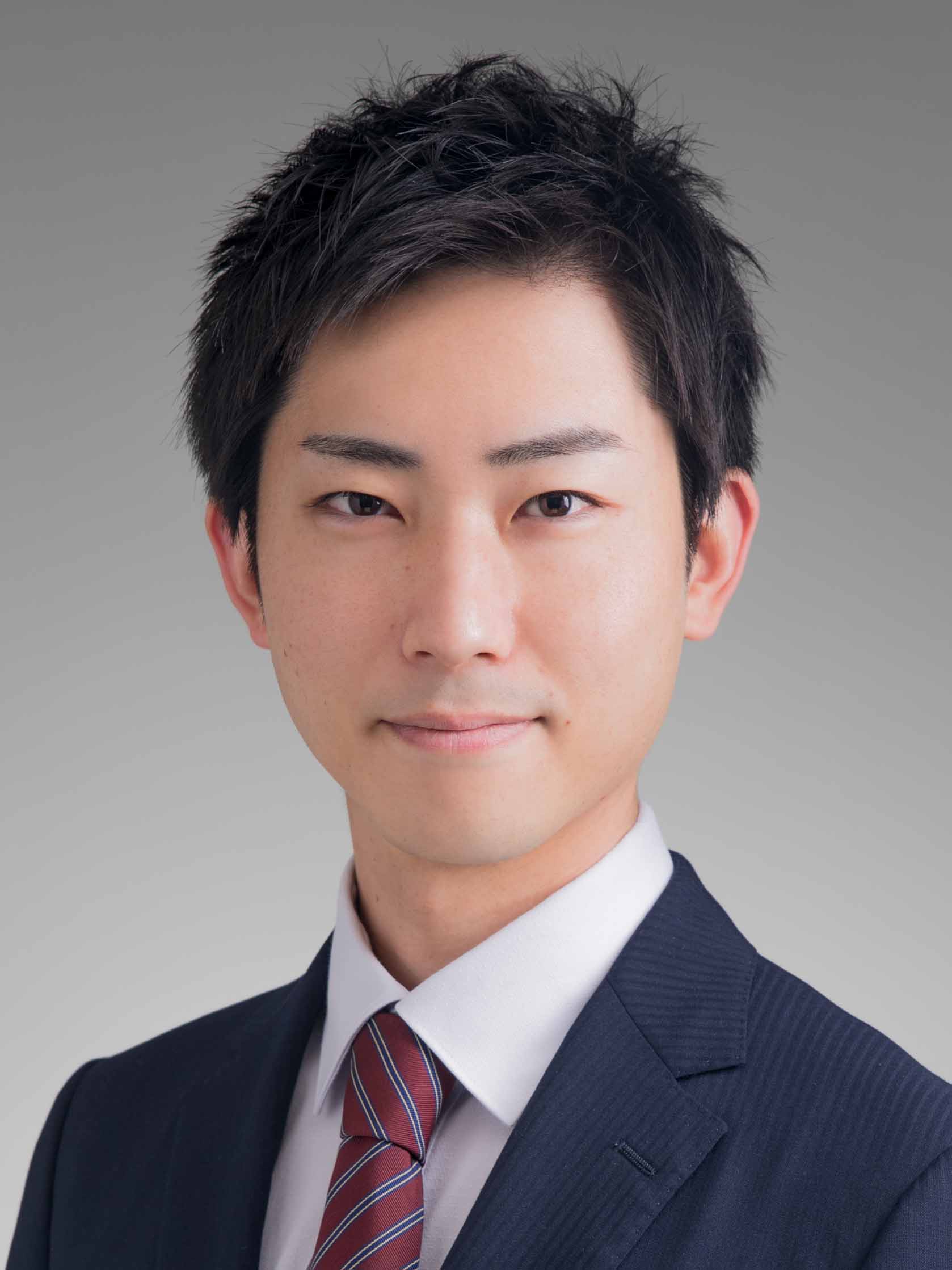}}]{Shuta Kikuchi} received the B.Eng. and M.Eng. degrees from the Waseda University in 2017 and 2019, respectively. He is currently pursuing a Ph.D. degree in applied physics at Keio University. His research interests include Ising machine, statistical mechanics, and quantum annealing. He is a member of the JPS.
\end{IEEEbiography}

\begin{IEEEbiography}[{\includegraphics[width=1in,height=1.25in,clip,keepaspectratio]{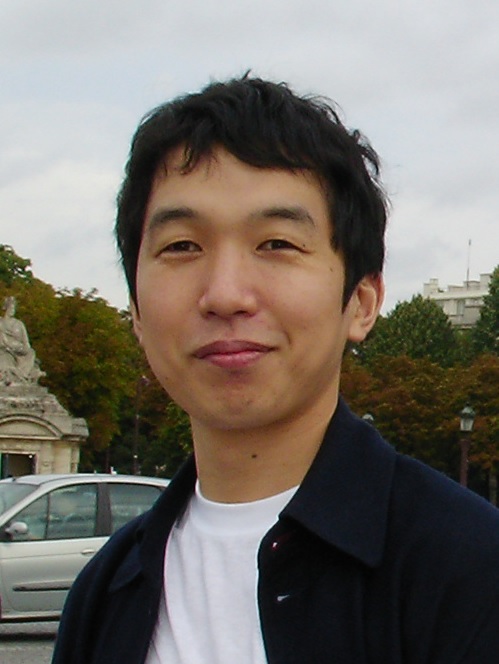}}]{Nozomu Togawa} (Member, IEEE) received the B.Eng., M.Eng., and Dr.Eng. degrees in electrical engineering from the Waseda University, Tokyo, Japan, in 1992, 1994, and 1997, respectively. He is currently a Professor with the Department of Computer Science and Communications Engineering, Waseda University. His research interests include quantum computation and integrated system design. He is a member of ACM, IEICE, and IPSJ.
\end{IEEEbiography}

\begin{IEEEbiography}[{\includegraphics[width=1in,height=1.25in,clip,keepaspectratio]{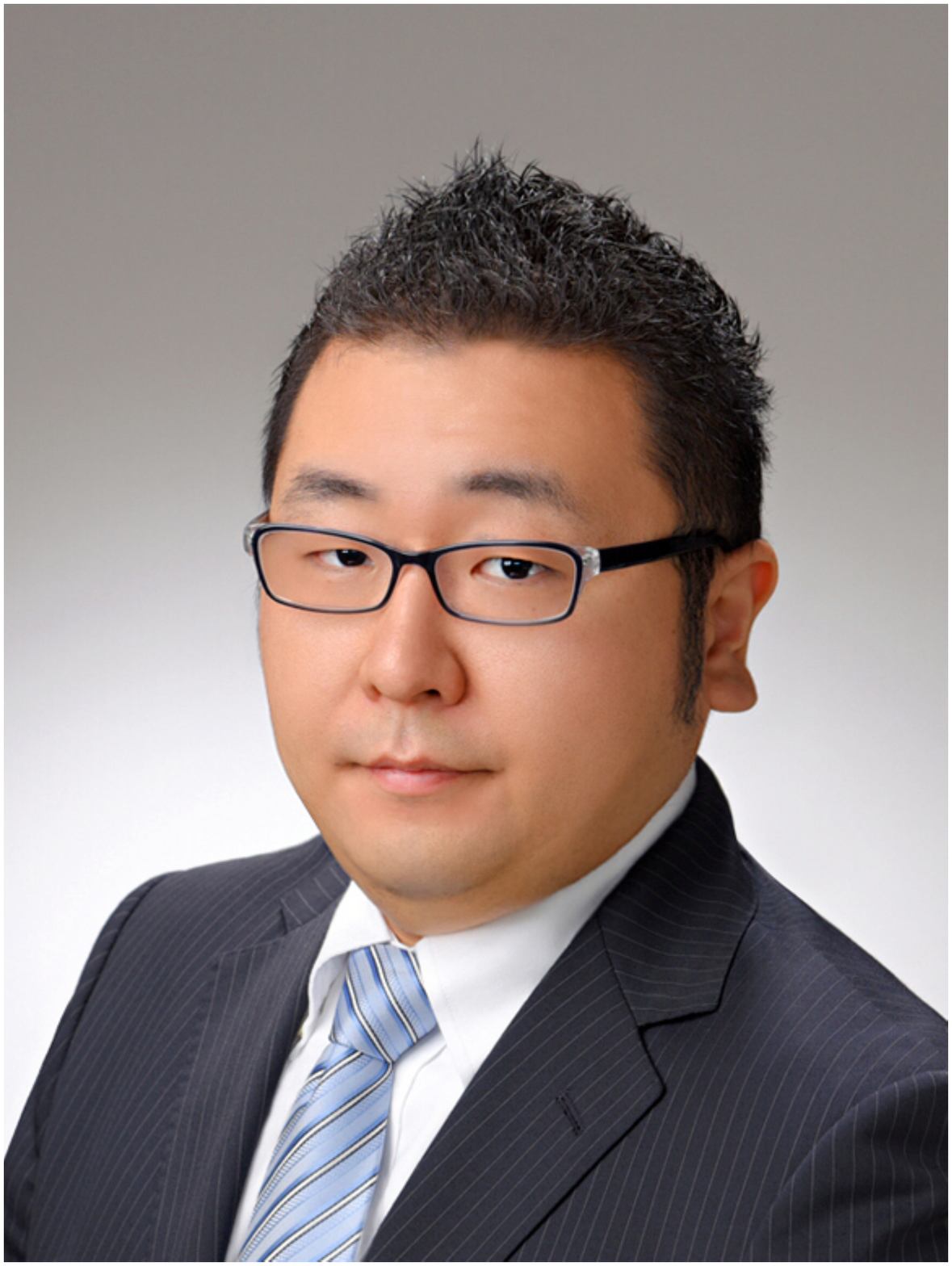}}]{Shu Tanaka} received the B.Sci. degree from the Tokyo Institute of Technology, in 2003, and the M.Sci. and Dr.Sci. degrees from The University of Tokyo, in 2005 and 2008, respectively. He is currently an Associate Professor with the Department of Applied Physics and Physico-Informatics, Keio University and a Core Director with Human Biology-Microbiome-Quantum Research Center (Bio2Q), Keio University. His research interests include quantum annealing, Ising machine, statistical mechanics, and materials science. He is a member of the JPS.
\end{IEEEbiography}

\EOD

\end{document}